\documentclass[a4paper,twocolumn,pra,superscriptaddress,amsmath,amssymb,mathrsfs,showpacs]{revtex4-1}
\usepackage{graphicx}
\usepackage{epstopdf}
\usepackage[table]{xcolor}     
\usepackage{braket}
\usepackage{algorithm}
\usepackage{algpseudocode}
\usepackage[colorlinks]{hyperref} 
\usepackage[caption=false]{subfig}

\hypersetup{%
linkcolor=blue,
citecolor=blue,
filecolor=black,
menucolor=black,
urlcolor=black
}

\begin{document}

\title{Maximizing Hole Coherence in Ultrafast Photoionization of Argon\\
with SPA-Optimization}

\author{R. Esteban Goetz}
\affiliation{Theoretische Physik, Universit\"{a}t Kassel,
Heinrich-Plett-Stra{\ss}e 40, D-34132 Kassel, Germany}

\author{Maximilian Merkel}
\affiliation{Theoretische Physik, Universit\"{a}t Kassel,
Heinrich-Plett-Stra{\ss}e 40, D-34132 Kassel, Germany}

\author{Antonia Karamatskou}
\affiliation{Center for Free-Electron Laser Science, DESY, Luruper
  Chaussee 149, D-22716   Hamburg, Germany}
  \affiliation{Department of Physics, Universit{\"a}t Hamburg,
    Jungiusstra{\ss}e 9, D-20355   Hamburg, Germany} 
    \affiliation{The Hamburg Centre for Ultrafast Imaging, Universit\"at Hamburg,
    Luruper Chaussee 149, D-22761 Hamburg, Germany} 

\author{Robin Santra}
\affiliation{Center for Free-Electron Laser Science, DESY, Luruper
  Chaussee 149, D-22716   Hamburg, Germany}
  \affiliation{Department of Physics, Universit{\"a}t Hamburg,
    Jungiusstra{\ss}e 9, D-20355   Hamburg, Germany} 
    \affiliation{The Hamburg Centre for Ultrafast Imaging, Universit\"at Hamburg,
    Luruper Chaussee 149, D-22761 Hamburg, Germany} 

\author{Christiane P. Koch}
\affiliation{Theoretische Physik, Universit\"{a}t Kassel,
Heinrich-Plett-Stra{\ss}e 40, D-34132 Kassel, Germany}
\email{christiane.koch@uni-kassel.de}

\date{\today}
\begin{abstract}
  Photoionization with attosecond pulses populates hole states in the photoion. 
  Superpositions of hole states represent ideal candidates for time-dependent spectroscopy, for example via pump-probe studies. 
 The challenge consists in identifying pulses that create coherent superpositions of hole states while satisfying practical constraints.
  Here, we employ quantum optimal control to maximize the degree of coherence between these hole states. To 
  this end, we introduce a derivative-free optimization method with Sequential
  PArametrization update (SPA-optimization). We demonstrate
  the versatility and computational efficiency of SPA-optimization for photoionization in argon 
  by maximizing the coherence between the $3s$ and $3p_0$ hole states using shaped attosecond
  pulses. We show that it is possible to maximize the hole coherence
  while simultaneously prescribing the ratio of the final hole state
  populations. 
\end{abstract}
\maketitle

\section{Introduction}

Quantum optimal control theory is a versatile tool for identifying 
external fields that steer the dynamics of a quantum system in a desired way~\cite{GlaserEPJD15}. Applications range from enhancing the signal-to-noise ratio in nuclear magnetic imaging to high-fidelity operations in quantum information science~\cite{GlaserEPJD15}. Besides the actual implementation of a desired task, unravelling the underlying control mechanism often serves a better understanding of the quantum system. This is referred to as quantum control spectroscopy. 

Photoionization is a prime tool for studying electron dynamics and electron correlations and as such it is a promising candidate for quantum control spectroscopy. Compared to other fields of application, quantum optimal control of photoionization is faced with two challenges. First, optimization algorithms have to be combined with time-dependent electronic structure methods. 
To date, this has been achieved for the time-dependent configuration interaction singles (TDCIS) method~\cite{KlamrothJCP06,GreenmanPRA15,GoetzPRA16}, 
the multi-configurational time-dependent Hartree-Fock
(MCTDHF) method~\cite{MundtNJP09} and time-dependent density functional
theory (TDDFT)~\cite{CastroPRL12,HellgrenPRA13}.
Second, the optimal control toolbox needs to be adapted to typical observables in photoionization processes. This includes, notably, photoelectron spectra and angular distributions. We have recently shown how
a complete 3D photoelectron spectrum or certain properties thereof can be targeted with quantum optimal control~\cite{GoetzPRA16}. 

In the present work, we shift the focus from controlling the photoelectron to controlling the
photoion. This is motivated by the progress in the observation of hole dynamics in the photoion~\cite{YoungPRL06,GoulielmakisNature10,WirthSci11,CalegariSci14}
which is initiated by the photoionization. A coherent superposition of hole states may be created through one-photon ionization by a pulse with sufficiently large bandwidth~\cite{PabstPRL2011} or through multi-photon processes~\cite{GoulielmakisNature10,WirthSci11}. 
Such  a superposition is the starting point for time-dependent spectroscopy 
of electron dynamics, for example via pump-probe studies to investigate hole alignment~\cite{YoungPRL06} or interchannel coupling~\cite{CalegariSci14}. As with any coherent spectroscopy, the degree of coherence of the state that will be transiently probed is a crucial resource~\cite{RybakPRL11}.  
However, the transient interaction between the photoion and the photoelectron introduces decoherence of the hole states even in one-photon ionization with attosecond pulses~\cite{PabstPRL2011}.  In optical tunnel ionization, the observed degree of coherence is also limited, so far to about 85 per cent~\cite{WirthSci11}. In that regime, even the shortest ionizing pulses do not allow to realize perfect coherence among the hole states~\cite{SantraPRA06}. Moreover, only outer-valence hole states are accessible and it is very hard to vary the population ratio of the hole states.

The challenge is thus to identify suitable pulses that create a desired superposition of hole states with predefined population ratio, satisfying practical constraints. This is the control problem that we consider here for the example of a superposition of the $3s$ and $3p_0$ hole states in the argon atom. Note that the 
$3s$ hole state in argon would be inaccessible in tunnel ionization. 
A necessary requirement for hole coherence is ionization into photoelectron states with the same angular momentum and energy. 
Because of the dipole selection rules, creating coherence between a pair of hole states through one-photon ionization may not be possible even if the spectral bandwidth of the ionizing pulse exceeds the energy separation of the two hole states. For multiphoton processes, it may be possible to generate hole coherence by ionization from occupied orbitals of opposite parity.
The use of quantum optimal control theory allows for exploring both regimes and,  moreover, for tackling the question of what the maximum degree of hole coherence is.

To this end, we employ a gradient-free optimization approach. It 
consists in choosing a suitable parametrization of the ionizing field and optimizing the parameters of the corresponding expansion. 
Importantly, we increase the number of optimization parameters sequentially as the optimization proceeds. 
This Sequential PArametrization update, or SPA-optimization, 
is key for ensuring sufficient flexibility in the representation of the field while avoiding the slow convergence that plagues gradient-free optimization for large numbers of optimization parameters. 

To actually carry out gradient-free optimization,  
numerous methods exist in the mathematics literature. However, quantum optimal control studies have so far used only the Nelder-Mead or downhill simplex method~\cite{DoriaPRL11,CanevaPRA11}. 
The standard Nelder-Mead approach is, however, prone to converge to local extrema, even for strictly convex functions~\cite{JeffreySIAM98}, 
which may lead to poor optimization results. 
Here, we compare this option for gradient-free optimization to the principal axis method, due to Brent~\cite{Brent1973}, and find the latter to be clearly superior both in terms of convergence speed and final value of the degree of coherence. 

The remainder of the paper is organized as follows. We present the theoretical framework in Sec.~\ref{sec:theory}, starting with the TDCIS equations in Sec.~\ref{subsec:tdcis}, defining the optimization problem in Sec.~\ref{subsec:control_target_dyanamics}, and outlining the optimization method in Secs.~\ref{subsec:optimization_methods} and \ref{subsec:sequential_update}.  Section~\ref{sec:optimization_results} is dedicated to a thorough numerical study of SPA-optimization. Taking as an example the maximization of coherence between the $3s$ and $3p_0$ hole states in argon, without any constraint on the respective hole populations, 
we illustrate the efficiency of the sequential parameter update, compare the Nelder-Mead to the principal axis method and demonstrate a significant speed-up of convergence due to a parameter scan prior to optimization. In Sec.~\ref{sec:prescribed_population}
we turn to the maximization of the hole coherence under the additional constraint of maintaining a certain population ratio for the hole states and study in depth the underlying control mechanism. Section~\ref{sec:concl} concludes. 

\section{Theoretical framework}
\label{sec:theory}

\subsection{Electron dynamics using TDCIS }
\label{subsec:tdcis}
We model the electron dynamics in photoionization 
by means of the time-dependent configuration interaction singles (TDCIS)
approach~\cite{GreenmanPRA10,xcid,PabstPRL2011,AntoniaPRA14}.  The TDCIS $N$-electron 
wavefunction reads
\begin{eqnarray}
  \label{eq:wavefunction}
  |\Psi(t)\rangle = \alpha_0(t)|\Phi_0\rangle
  + \sum_{i,a}\alpha_i^a(t)|\Phi_i^a\rangle\,,
\end{eqnarray}
where $|\Phi_0\rangle$ and $|\Phi_i^a\rangle$ denote the 
Hartree-Fock ground state and the single particle-hole excitation 
from an initially occupied orbital, labeled $i$, to an initially unoccupied  orbital $a$.
The binding energies utilized in the present work are those obtained from the Hartree-Fock formalism using Koopmans'
theorem~\cite{Koopmans1934,Grand2007relativistic}. The dynamics is governed by the time-dependent Hamiltonian,
\begin{eqnarray}\label{eq:H}
  \hat{H}(t) = \hat{H}_0 + \hat{H}_1 + E(t)\hat{z}\,,
\end{eqnarray}
where $\hat{H}_0$ is the mean-field Fock operator and $\hat{H}_1$ is the residual
Coulomb interaction,
\begin{eqnarray}
  \hat{H}_1 = \hat{V}_C - \hat{V}_{MF}\,, 
\end{eqnarray}
with $\hat{V}_C$ and $\hat{V}_{MF}$ being the electron-electron interaction and the
mean-field potential, respectively. The last term on the right-hand side of Eq.~\eqref{eq:H} describes the electric dipole interaction of the atom with an external electric field, assumed to be linearly polarized.

The photoion corresponds to a reduced system that is obtained by integrating out the photoelectron and thus needs to be described by a density matrix~\cite{SantraPRA06}. 
To study the hole dynamics, we use the ion density matrix approach of Refs.~\cite{PabstPRL2011,GreenmanPRA10},
\begin{eqnarray}
\label{eq:ion_density_matrix}
\rho^{IDM}_{i,j}(t) = \text{Tr}_a\left[ |\Psi(t)\rangle\langle\Psi(t)| \right]_{i,j}
= \sum_{a}\langle\Phi_i^a|\Psi(t)\rangle\langle\Psi(t)|\Phi_j^a\rangle\,,\nonumber\\ 
\end{eqnarray}
where the trace is carried out over the virtual channels which are occupied by
the photoelectron. 
In order to avoid numerical artifacts due to  reflection on the edges of the numerical grid as the TDCIS wavefunction propagates over time, a complex absorbing potential (CAP)~\cite{SantraPhysRep02,MugaPhysRep04} of the form
\begin{eqnarray}
\label{eq:CAP}
-i\eta\hat{W}(\hat{r}) = -i\eta h(\hat{r}-r_c)\times \left(\hat{r}-r_c \right)^2 
\end{eqnarray}  
is utilized~\cite{RissJPB93,JolicardChemPhysLett85,GreenmanPRA10,RohringerPRA09}. In Eq.~\eqref{eq:CAP}, $h(\cdot)$, $r$ and $r_c$ refer to the Heavyside distribution, the
distance from the origin and the critical distance at which the CAP starts
absorbing, respectively. The CAP affects all virtual orbitals and thus also the ion density matrix, which therefore must be corrected according to~\cite{GreenmanPRA10,RohringerPRA09}
\begin{subequations}
\begin{eqnarray}
\label{eq:correction}
\rho^{IDM}_{i,j}(t) &=&\tilde{\rho}^{IDM}_{i,j}(t) + 2\eta\,
e^{(\varepsilon_i-\varepsilon_j)t}\nonumber\\
&&  \times\sum_{a,b}w_{a,b}\int^{t}_{-\infty}dt^\prime\,
\alpha^a_{i}(t^\prime)\alpha^{*b}_{j}(t^\prime)e^{(\varepsilon_i-\varepsilon_j)t^\prime}\,,\quad\quad 
\end{eqnarray}
where the ``uncorrected'' matrix elements of ion density matrix $\tilde{\rho}^{IDM}(t)$ read~\cite{GreenmanPRA10,RohringerPRA09}

\begin{eqnarray}
  \tilde{\rho}^{IDM}_{i,j}(t) &=&
  \sum_{a} (\Phi^a_i|\Psi(t)\rangle\langle\Psi(t)|\Phi^a_j)\,, 
  \label{eq:uncorrected_rho_ij}
\end{eqnarray}
\end{subequations}
with $|\Phi^a_j) =|\Phi^a_j\rangle $ and $|\Phi^a_j\rangle$ and $(\Phi^a_j|$ referring to the right and left eigenvectors of $\hat{F}-i\eta\hat{W}$, where $\hat{F}$ is the Fock operator.
Note that, due to the CAP, $(\Phi^a_j|$ and $|\Phi^a_j\rangle$ are not orthogonal~\cite{GreenmanPRA10}.

Equation~\eqref{eq:correction} provides the starting point for
defining a measure of hole coherence: The positive
semidefinite quantity
\begin{eqnarray}
  \label{eq:tildeg_ij}
  g_{i,j}(t) = \dfrac{|\rho^{IDM}_{i,j}(t)| }{\sqrt{\rho^{IDM}_{i,i}(t) \rho^{IDM}_{j,j}(t)}}
\end{eqnarray}
defines the degree of coherence between the hole states in the atomic
orbitals $i$ and $j$~\cite{PabstPRL2011}. For a totally incoherent
statistical  mixture $g_{i,j}=0$ , whereas $g_{i,j}=1$
for perfect coherence between the states $i$ and $j$. 

We will analyze below the impact of the Coulomb interaction on the
hole coherence. To this end, we will compare the ``full'' (or
interchannel) model and the intrachannel approximation. Within the
``full'' model, the photoelectron may couple to all hole states in the
parent ion which mediates a coupling between different channels. In contrast, 
within the intrachannel approximation, the photoelectron can only
interact with the hole in the orbital from which 
it originates~\cite{PabstPRL2011}.

Moreover, it will be useful to quantify how fast a photoelectron leaves the parent ion. 
To this end, we can exploit that the CAP acts as a sensor for the excited electron, or eventually, the photoelectron 
to reach the asymptotic region, where the CAP is active. Such an indicator is given by
\begin{eqnarray}
\label{eq:Delta}
\Delta_{\rho}(t) =1- \Big(\text{Tr}_{i}\left[\tilde{\rho}^{IDM}(t)\right] + |\alpha_0(t)|^2\Big),
\end{eqnarray}
since $\text{Tr}_i[\tilde{\rho}^{IDM}(t)] +
  |\alpha_0(t)|^2 $ is not equal to one, due to the CAP
(only $\text{Tr}_i[\rho^{IDM}(t)] + |\alpha_0(t)|^2 $ is)
and the CAP does not affect the
coefficients $\alpha_0(t)$.

\subsection{Optimization problem}
\label{subsec:control_target_dyanamics}

Our optimization targets maximization of hole coherence. In a first stage, we 
maximize the degree of coherence
between the $3s$ and $3p_0$ hole states in argon at the final time $T$, 
regardless of the final hole population ratio in the $3s$ and $3p_0$
orbitals. It is customary to minimize rather than maximize, such that 
the final-time cost functional reads
\begin{eqnarray}
 J_T^{(1)} &=&  
 \left( g_{3s,3p_0}(T)-1\right) ^2 \,.
 \label{eq:functionalf0}
\end{eqnarray}
It takes values between 0 and 1 with $J^{(1)}_T=0$ corresponding to
perfectly coherent $3s$ and $3p_0$ hole states. 

When the target is not only to maximize hole coherence but also to
prescribe a certain ratio $\mathcal R$ between the hole populations, 
the final time cost functional becomes 
\begin{eqnarray}  \label{eq:functionalf}
  J_T^{(2)} &=&
  w_{pop}\left(\dfrac{\rho_{3p_0,3p_0}(T)}{\rho_{3s,3s}(T)} - \mathcal
  R\right)^2 \\ && + 
 w_{coh}\left( g_{3s,3p_0}(T)-1\right) ^2\,,\nonumber
\end{eqnarray}
where $w_{pop}$ and $w_{coh}$ are optimization weights that can be
used to stress the relative
importance of each term in Eq.~\eqref{eq:functionalf}. 

Additional constraints in functional form, that are customary in 
gradient-based optimization and often cumbersome to implement~\cite{PalaoPRA13,ReichJMO14}, 
are not needed when using
gradient-free optimization: The bandwidth of the field is determined by
the allowed frequency range, and the maximal amplitudes of the Fourier
components can be directly confined by choice of sampling range.

\subsection{Optimization method}
\label{subsec:optimization_methods}

We opt here for gradient-free optimization which only requires evaluation
of the functional but not its gradient. This avoids backward
propagation of an adjoint state that is typical for gradient-based
optimization approaches~\cite{ReichJCP12}. In our case, backward
propagation involves an inhomogeneous Schr\"odinger equation with the
inhomogeneity originating from the correction of the ion density
matrix due to the presence of the CAP,
cf. Eq.~\eqref{eq:correction}. While a numerically exact solution of
inhomogeneous Schr\"odinger equations is possible~\cite{NdongJCP09},
it becomes challenging if the source term gets large. This is the case
here. 

A number of methods for gradient-free optimization exists. A popular
approach, and notably the only one employed in quantum optimal control
so far~\cite{DoriaPRL11,CanevaPRA11}, is due to Nelder and
Mead~\cite{NelderMead}. It minimizes a function of $n$
  optimization parameters (therefore gradient-free approaches are
  sometimes referred to as
  parameter optimization) by
comparing function evaluations at the $n+1$ vertices of a general
simplex, and updating the worst vertex by moving it around a new
vertex that is an average of the remaining (best) vertices~\cite{Klein2013,NelderMead}. While the approach often works
well, it may become ill-conditioned, particularly when non-convex
forms of the function are involved.
As an alternative to the Nelder-Mead simplex approach, we consider the principal axis optimization method~\cite{Brent1973} which is based on an inverse parabolic interpolation. 

The advantage of avoiding backward propagation of the adjoint state
with gradient-free optimization is balanced by two drawbacks---the
requirement for prior parametrization of the field, and the
convergence not being monotonic. Gradient-free optimization may 
lead to poor fidelities if (i) the parametrization of the field
is not properly chosen, (ii) the number of parameters is too small,
or, paradoxically, (iii) the number of parameters exceeds a
certain threshold. In the latter case, a saturation effect causes the
functional to reach an asymptote very quickly and the optimization
gets stuck. In order to circumvent this problem, we employ
a sequential parametrization update technique
which is explained in the following.

\subsection{Sequential optimization update}
\label{subsec:sequential_update}

The poor performance of gradient-free optimization due to a
too large number of optimization parameters can be avoided by 
a sequential update of the number of optimization parameters~\cite{RachPRA15}. 
Here, we adopt this approach to optimization methods beyond a
Nelder-Mead simplex search and allow for treating the circular frequencies
themselves as optimization parameters while still maintaining a
prespecified bandwidth. The optimization is started with
a minimal number of parameters, and additional parameters are included
\textit{on-the-fly} as the algorithm proceeds iteratively, i.e., every time
the value of the optimization functional reaches
a plateau. 

As an example of the SPA technique, consider parametrization of the field by Fourier components, 
\begin{eqnarray}
\label{eq:field_param}
E_{NI}(t) &=&  \sum^N_{n=1} 
\sum_{i=1}^{I} s_n(t,\sigma_n)\,\big\{
 f_n(a_{n,i})\cos(\omega_{n,i}\, t) \\
  &&\hspace*{16ex}+ f_n(b_{n,i})\sin(\omega_{n,i}\,t)\big\}\,,\nonumber
\end{eqnarray}
with the Fourier amplitudes $a_{n,i}$ and $b_{n,i}$ as optimization
parameters. The double sum notation was chosen to ease implementation
of a field that consists of $N$ subpulses. 
In Eq.~\eqref{eq:field_param}, $s_n(t,\sigma_n)$ is a fixed envelope, for example Gaussian or $\sin^2$-shaped. 
The durations $\sigma_n$ of the subpulses as well as the circular frequencies $\omega_{n,i}$ can be fixed or considered as additional optimization parameters. The functions $f_n(\cdot)$ are introduced in order to constrain the Fourier amplitudes $a_{n,i}$ and $b_{n,i}$ to within
a prespecified range. For instance, a function of the form
\begin{subequations}
  \label{eq:transformations}
  \begin{eqnarray}
  \label{eq:transformation1}
  f_n(\zeta_i) = \zeta_o\,  \int^{\zeta_i}_0 e^{-t^2}\,dt
  \label{error}
\end{eqnarray}
ensures that the Fourier coefficient does not exceed a given maximum absolute value $\zeta_o$, avoiding large amplitudes for the resulting optimized field. Equivalently, a hyperbolic tangent form, 
\begin{eqnarray}
  \label{eq:transformation2}
  f_n(\zeta_i) = \zeta_{n,o}\,  \dfrac{e^{\zeta_i} - e^{-\zeta_i}} {e^{\zeta_i} + e^{-\zeta_i}}\, ,
  \label{eq:tangent}
\end{eqnarray}
\end{subequations}
may be utilized to control the maximal amplitude of the optimized
field. 
One could also apply the transformations~\eqref{eq:transformations}
  to the overall
  electric field instead of each Fourier component separately. This
  may, however, result in low frequency components. Such artifact
  frequencies are undesirable, in particular when the solution shall be constrained to a given spectral range. 

To start the optimization, we  choose, for simplicity, a single pulse, $N=1$, with two Fourier amplitudes, $I=2$, and fixed or variable circular frequencies. 
When using fixed circular frequencies, a set of circular frequency values is specified in the very beginning, which are successively added during the parametrization updates. If the circular frequencies are treated as optimization parameters, 
the spectral range can be controlled by restricting the circular frequencies to an interval via the mapping 
\begin{eqnarray}
 \label{eq:freq_map}
 \omega^{new} = \frac{1}{2}(\omega_{max} - \omega_{min})\tanh(\omega)
  + \frac{1}{2}\left(\omega_{max} + \omega_{min} \right)\nonumber\,,\\
\end{eqnarray}
where $\omega\in \rm I\!R $ is the circular frequency
  returned by the optimization algorithm, whereas 
$\omega^{new}$, which is guaranteed be in the interval
$]\omega_{min},\omega_{max}[$ by Eq.~\eqref{eq:freq_map}, is the one used for the propagation.

Consider for simplicity the example of fixed circular frequencies, treating  
the pulse duration (full width at half maximum (FWHM) of the intensity
profile), Fourier amplitudes and relative phases as optimization
parameters. The procedure consists of two loops, an outer loop over
generations (with each generation corresponding to a parametrization
with $m$ parameters), and an inner loop, iterating for a given
parametrization. The inner loop proceeds until $N_c$ evaluations of
the functional, i.e. propagations of the wavefunction, are reached. It
then checks whether the overall minimization threshold is reached. If
so, the complete procedure is stopped; if not, it checks whether the
value of the functional has changed 
significantly during the $N_c$ iterations. If so, another $N_c$
iterations are carried out, if not, then the algorithm
increases  the number of optimization parameters, and restarts the
optimization for the new generation, using the best previous field as
guess field for the new parametrization with all new optimization
parameters set to zero. 
This procedure of updating the parametrization of the field is
repeated every time that the functional gets stuck, allowing it to
escape from  the plateau.
The user needs to specify the maximal number of generations $\mathcal G_{max}$, or new parametrizations, together with $N_c$, the maximum number of evaluations of the functional, i.e. propagations, and the tolerance thresholds. 

In the following, we show that such a sequential parametrization update is  more efficient than choosing a large number of parameters from the beginning. In a sense, the optimization is ``driven''
efficiently and does not get stuck in a final plateau since 
every time the functional reaches a saturation plateau, the additional
parameters introduced allow for escaping from such an asymptotic
region. This is in line with the findings of Ref.~\cite{RachPRA15}
where the frequencies are randomized within a prespecified interval. 
Furthermore, we show that updating the parametrization is
particularly efficient when combined with the principal
axis method, due to Brent~\cite{Brent1973}, as compared to the
Nelder-Mead optimization algorithm~\cite{NelderMead}, employed in
Ref.~\cite{RachPRA15}.

\section{Maximization of hole coherence with arbitrary population ratio}
\label{sec:optimization_results}

\subsection{Numerical performance of SPA-optimization}
\label{subsec:arbitrary}
 
The goal is to maximize the degree of coherence $g_{i,j}(T)$ between the $3s$ and $3p_0$ hole-population in argon, using an  electric field, linearly  polarized along the $z$ direction, in the XUV regime with the maximal field amplitude not exceeding $0.02\,$a.u. Correspondingly, the target functional is the one defined in Eq.~\eqref{eq:functionalf0}.
The wavepacket is represented, according to Eq.~\eqref{eq:wavefunction}, in 
terms of the ground state $|\Phi_0\rangle$ and 
excitations $|\Phi^a_i\rangle$, from which the corrected form of the IDM, due
to the CAP, cf. Eq.~\eqref{eq:correction}, is calculated.
The  calculations employed a pseudo-spectral grid with
density parameter $\zeta=0.50$~\cite{GreenmanPRA10}, a  
spatial extension of $200\,$a.u. 
and 800 grid points, with angular momentum functions restricted to
$L_{max} = 10$.   
A CAP strength $\eta$ in Eq.~\eqref{eq:CAP}, $\eta = 0.002$, and absorbing
radius $r_c$ in Eq.~\eqref{eq:CAP}, $r_{c} =
180.0\,$a.u., are chosen.

\begin{figure}[tb]
\centering
\includegraphics[width=0.95\linewidth]{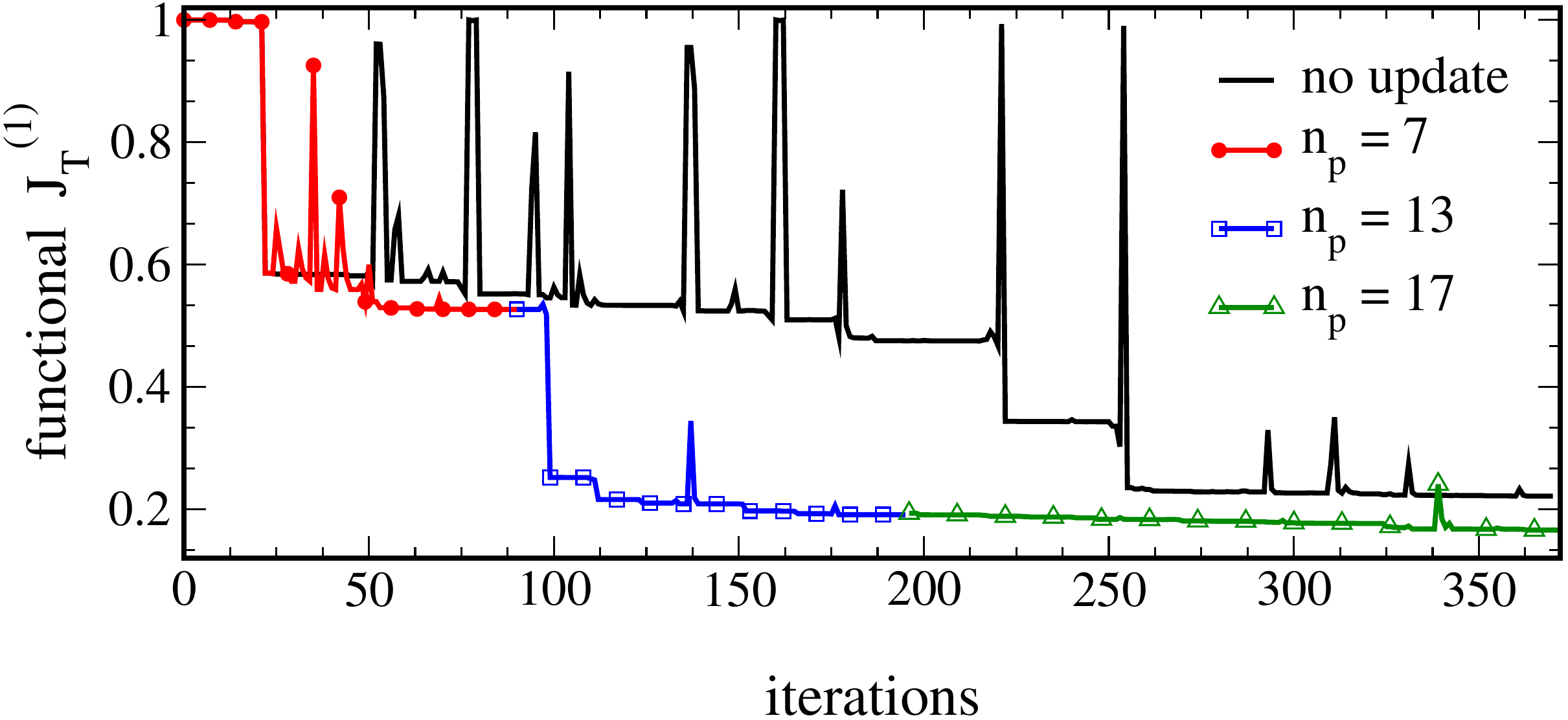}
  \caption{
    Efficiency of optimization using the principal axis method of
    Brent with fixed parametrization (black line, $n_p=13$) and with sequential
    parametrization update (SPA-optimization, colored lines). 
    SPA-optimization converges significantly faster and yields a
    better hole coherence.
  }
\label{fig:plot1a}
\end{figure}
We first compare  our sequential parametrization update (SPA) technique to 
optimization with a standard fixed parametrization, using the principal axis method in both cases to determine the change in parameters.
Figure~\ref{fig:plot1a} shows the optimization efficiency for the two methods, started with the same guess field.
The optimization parameters are the pulse duration and the Fourier
components. The circular frequencies, taken to be fixed on an evenly spaced frequency grid, are chosen in the XUV regime. For the standard version, the entire
frequency grid is used from the beginning of the optimization, while for 
SPA-optimization circular frequencies from the grid are successively added.
The standard non-updated version (full black line), for which the field is defined by $13$ optimization parameters, decreases quasi-monotonically but very slowly during the first $210$ iterations. Then the 
functional considerably decreases between the iterations 210 to 250
before reaching a plateau with  final value  $J^{(1)}_T = 0.21$.
SPA-optimization is started by defining at first a pulse characterized by $7$ circular frequencies, which coincide with the first seven circular frequencies
from the overall set of circular frequencies. 
After $50$ iterations with these parameters, SPA-optimization reaches already a functional value slightly below that
reached by the non-sequential version after the same number of iterations. 
Once the plateau for the field containing $7$ optimization parameters is
reached, the new generation is started by adding $6$ additional optimization
parameters. As can be seen from Fig.~\ref{fig:plot1a}, such an update allows 
the functional to considerably decrease, reaching after just $100$ iterations 
the same value that is 
obtained with the non-sequential version in $255$ iterations. Furthermore, it
also shows that there are some frequency components resulting from the
non-update version, that are not  necessarily required for the optimization. 
The different colors in Fig.~\ref{fig:plot1a} illustrate the increase in the number of optimization parameters as a function of the number of propagations. 
From Fig.~\ref{fig:plot1a}, it is clear that the sequential parametrization update version is more efficient than standard optimization: It allows not only to reach higher fidelities at the end of the optimization, but also  converges faster.
The comparison shown in Fig.~\ref{fig:plot1a} does not depend on the specific choice of the initial guess. That is, we have carried out the comparison for several guess fields and observed always a better performance of SPA-optimization compared  to optimization with fixed parametrization.

\begin{figure}[tb]
\centering
\includegraphics[width=0.95\linewidth]{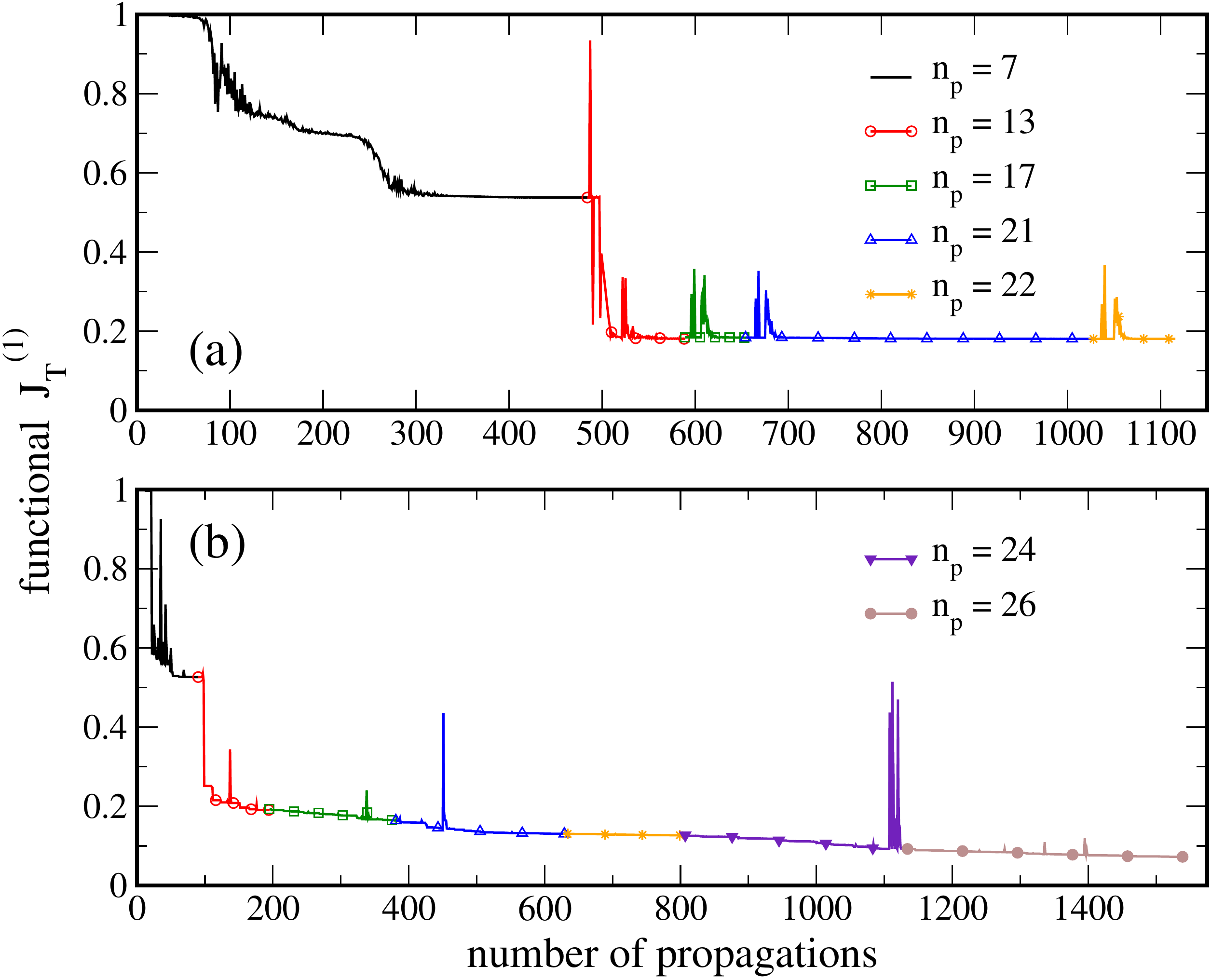}
\caption{
  Comparison of SPA-optimization using (a) Nelder-Mead simplex search
  and (b) the principal axis method of Brent. The same initial guess
  field was utilized in both cases. SPA-optimization with the
  principal axis method converges significantly faster and yields a
  better hole coherence than with the Nelder-Mead simplex search. 
}
\label{fig:plot2}
\end{figure}
It is clear that the SPA-approach can be extended to other
gradient-free optimization methods. A particularly popular method is 
the widely used Nelder-Mead downhill simplex approach, which we now compare to  the principal axis method. The  convergence behavior of the two methods, when using the SPA-technique, is shown in Fig.~\ref{fig:plot2}. Both Nelder-Mead simplex and principal axis method are again started with 7 parameters, as described above, using the same guess 
for both methods. The principal axis method is found to clearly outperform the Nelder-Mead simplex: Indeed, with only $7$ optimization parameters, the principal axis method reaches a value of $J_T^{(1)}=0.50$ already after $100$ iterations, whereas the simplex method requires almost $400$ iterations to reach the same value. Moreover, the simplex algorithm tends to reach a plateau more easily than the principal axis method,
and after 600 iterations, the functional does not decrease even upon increasing the number of parameters.
This behavior is typical, and we only show representative
results in Fig.~\ref{fig:plot2}. For example, changing the number of critical iterations does not change this observation---the
Nelder-Mead simplex method tends to get stuck more rapidly and the
optimization cannot escape from the plateau, cf. the blue triangles in  Fig.~\ref{fig:plot2}(a). In contrast, as seen from  Fig.~\ref{fig:plot2}(b), with the principal axis method the functional continues to decrease, albeit slowly, when the number of optimization parameters is increased. According to our numerical experiments, this behavior is again  independent of the guess field.

\begin{figure}[tb]
\centering
\includegraphics[width=\linewidth]{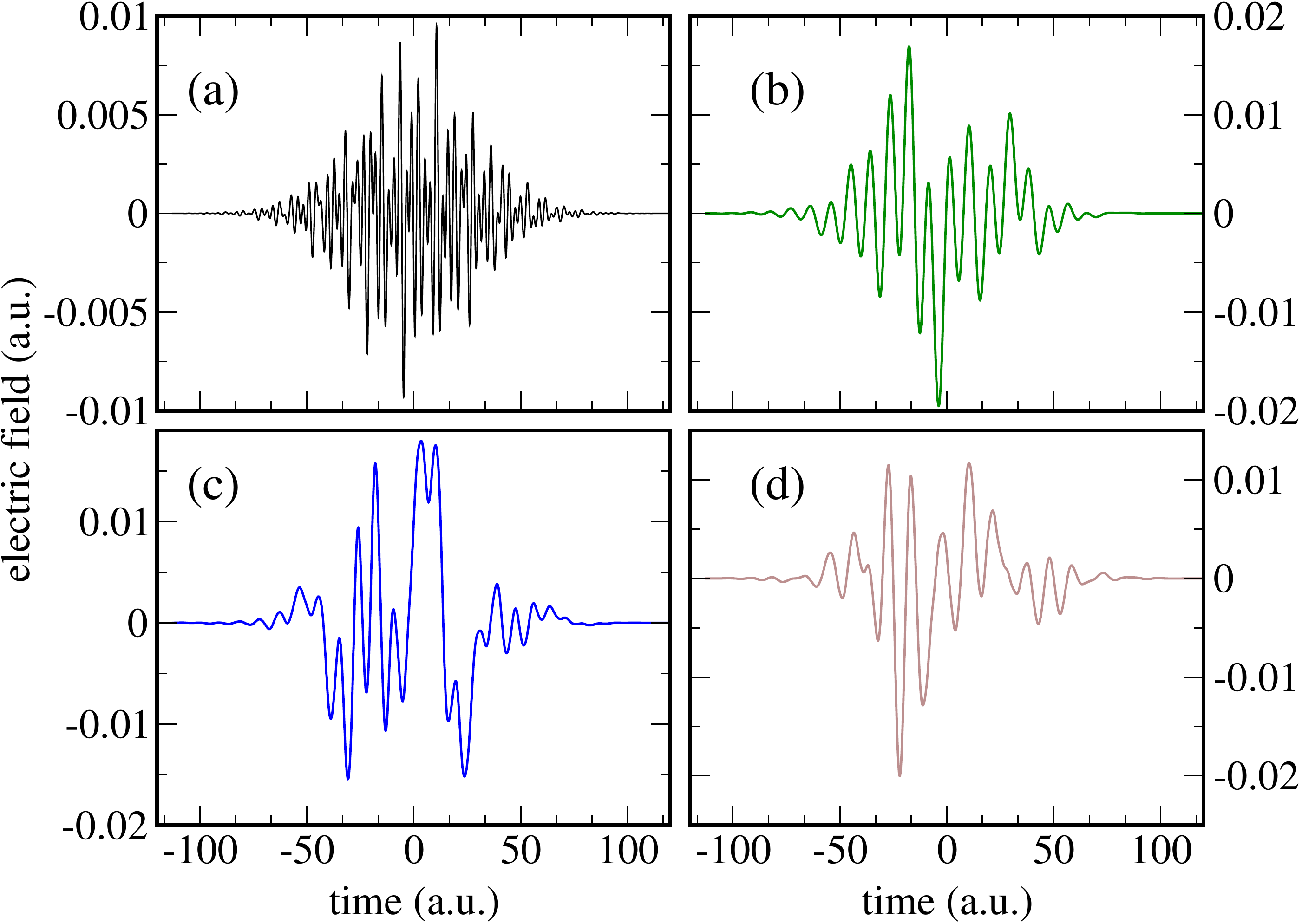}
\caption{Optimized fields obtained with SPA-optimization using the
  principal axis method (same color code as in
  Fig.~\ref{fig:plot2}). The field with 26 parameters, 
  shown in panel (d), yields a 
  degree of coherence of $g_{3s,3p_0}(T)=0.989$. 
}
\label{fig:plot3}
\end{figure}

We thus find that SPA-optimization based on the principal axis method  represents a promising alternative not only to the widely used Nelder-Mead simplex approach, but also to the principal axis method itself, when used in the standard version with a fixed number of optimization parameters. For completeness, we present in Fig.~\ref{fig:plot3} the optimized fields found at the different stages of the update procedure, using the same color code as in 
Fig.~\ref{fig:plot2}. Comparison of Figs.~\ref{fig:plot3}(b) and (c) with Fig.~\ref{fig:plot2}(b) shows that, although both fields have very different shapes and maximal amplitudes, they lead to similar hole coherences, 0.56 and 0.55, respectively. 
The final optimized field is  depicted in Fig.~\ref{fig:plot3}(d). 
Its frequency components lie in the XUV regime by construction and the maximal field amplitude is constrained to below 0.02$\,$a.u. as desired. The resulting degree of coherence amounts to 
$g_{3s_0,3p_0}=0.989$ after 1500 iterations. 

All numerical experiments that we have carried out reproduced the relative advantage of SPA-optimization over optimization with a fixed number of optimization parameters (Fig.~\ref{fig:plot1a}) and of the principal axis method over Nelder-Mead simplex (Fig.~\ref{fig:plot2}). However, they also revealed a rather high sensitivity of the optimization success, both in terms of convergence speed and final hole coherence achieved, on the initial guess. This suggests to pre-scan the parameters of the initial guess,  as studied next.

\subsection{Optimization using a ``pre-optimized''  guess field}
\label{subsec:brute_force}

The idea is to identify a small number of key parameters whose values are scanned in a prespecified range. While this does not constitute optimization in itself, 
it is related in spirit to the hybrid optimization approach of Ref.~\cite{GoerzEPJQT15} which combines a 
cheap, low-level parameter ``pre-optimization''  with a numerically more expensive, high-level gradient-based optimization. Once the parameter 
scan has been carried out, the best parameters resulting from the scanning procedure, i.e the ones that minimize, at least locally, 
the functional of interest, are chosen to define the guess for the actual SPA-optimization.
As a result, the actual optimization is started with a minimal number of
optimization parameters at an already relatively good fidelity. 

This approach is particularly useful when no a priori physical insight into the best choice of the field parameters is available. The required calculations are independent of each other and can thus be carried out in parallel. Nevertheless, the number of parameters to be scanned should be kept at a minimum. Furthermore, it is not necessary to perform the scan with very high resolution since small changes in the parameters that  significantly improve the target will be readily identified by the subsequent optimization.

\begin{figure}[tbp]
\centering
\vspace{-0.1cm}
\subfloat{
\includegraphics[clip,width=0.95\columnwidth]{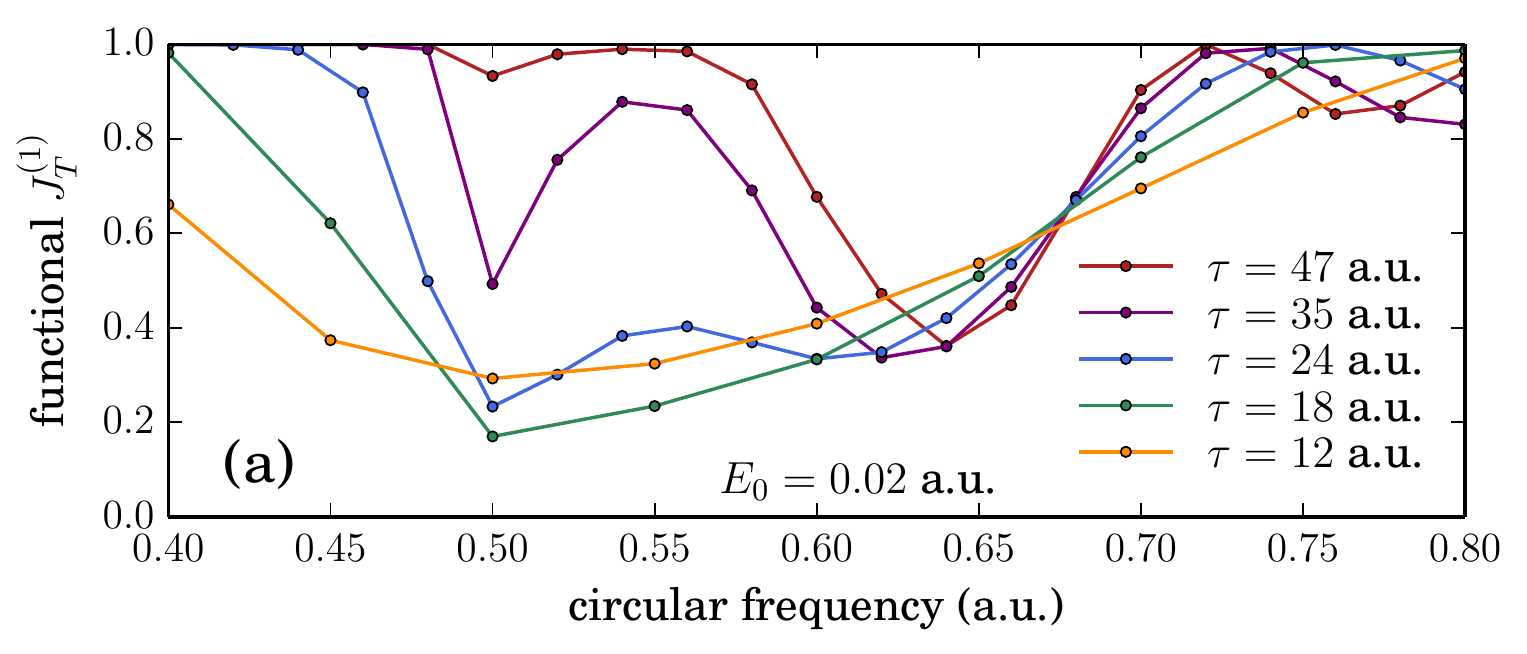}
}
\vspace{-0.4cm}
\subfloat{
\includegraphics[clip,width=0.95\columnwidth]{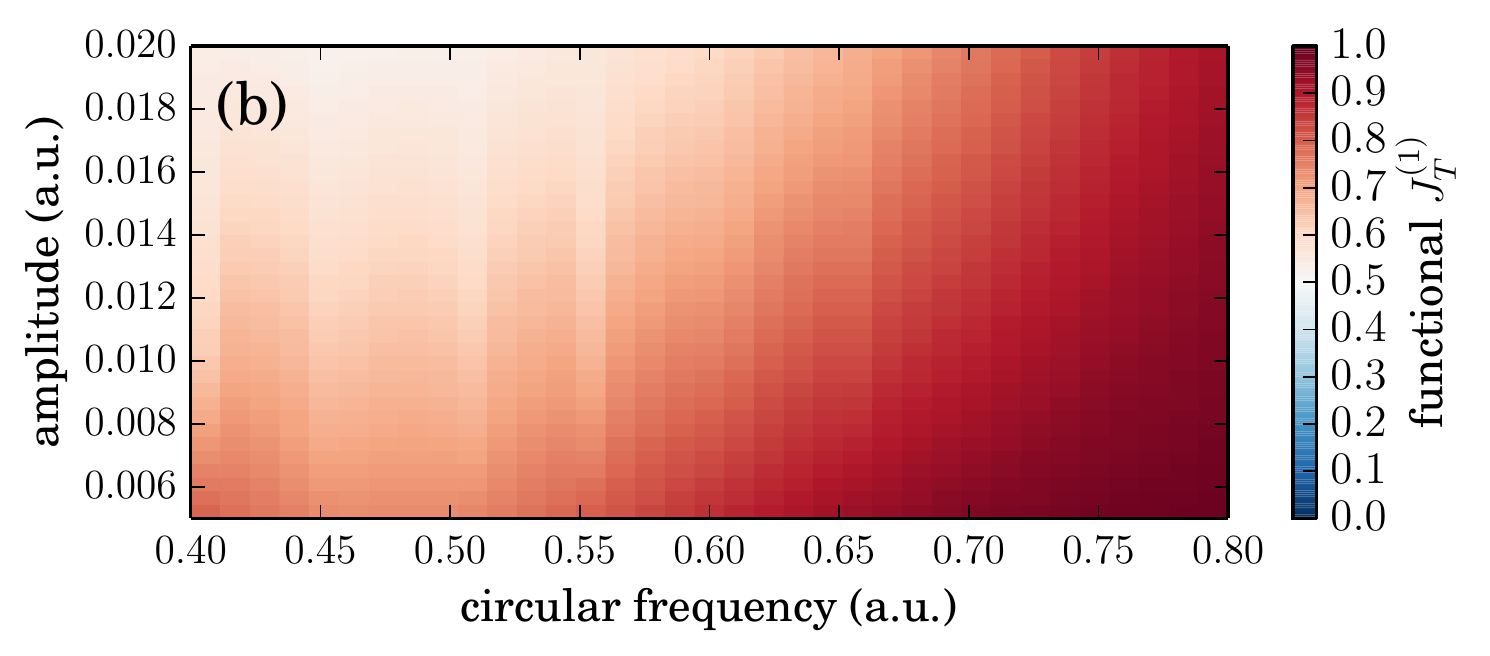}
}
\vspace{-0.4cm}
\subfloat{
\includegraphics[clip,width=0.95\columnwidth]{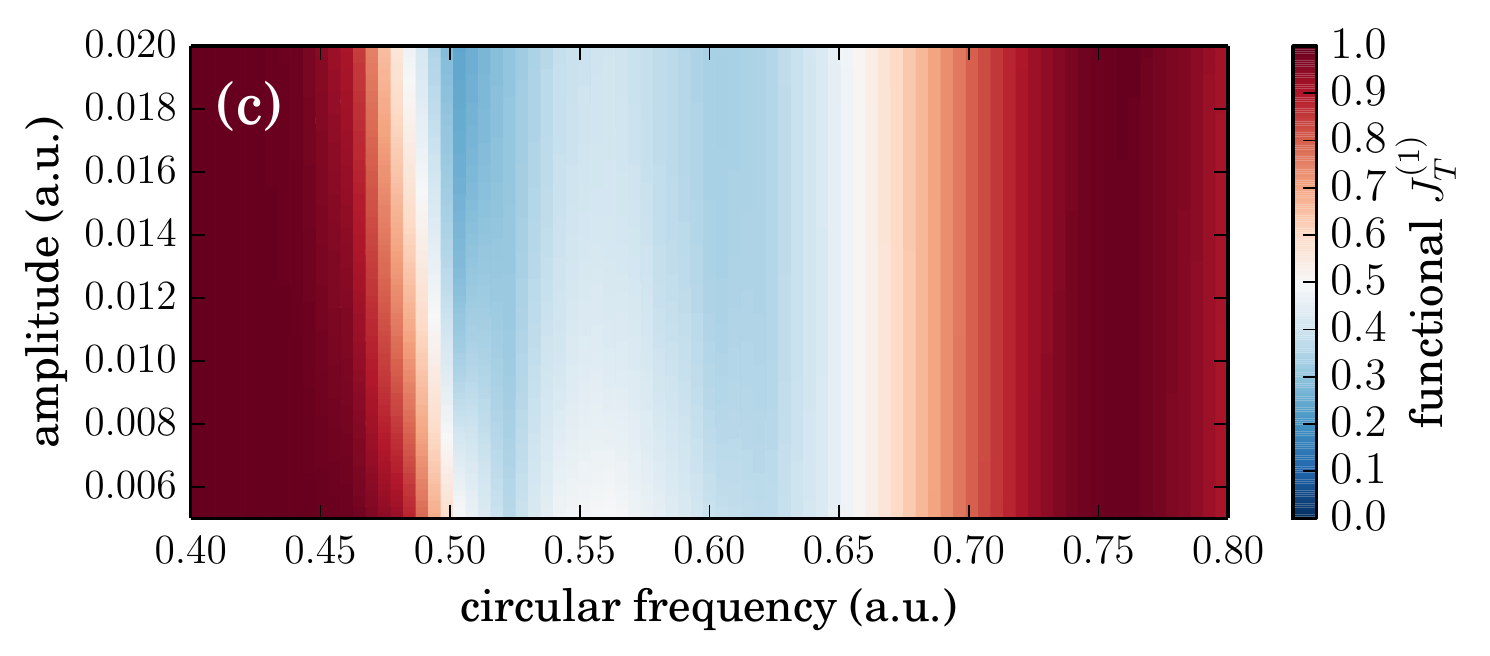}
}
\caption{Parameter scanning prior to optimization:
  Frequency scan with fixed peak amplitude for several pulse durations (FWHM of
  the intensity)~(a) and joint frequency / peak amplitude scans for fixed pulse
  durations $\tau=6\,$a.u.~(b) and $\tau=23\,$a.u.~(c).
  Favorable parameters for the initial guess field can clearly be
  identified. 
}
\label{fig:plot4}
\end{figure}
We scan in the following three parameters of a transform-limited Gaussian pulse---its peak amplitude, central frequency, and duration or, equivalently, spectral width. The results are shown in 
Fig.~\ref{fig:plot4}. Keeping the peak amplitude fixed at $E_0=0.02\,$a.u. and varying the pulse frequency, one broad minimum of the functional is observed in Fig.~\ref{fig:plot4} (top)
for short (spectrally broad) pulses near $\omega_{ph}=0.50\,$a.u. This minimum is shifted to $\omega_{ph}=0.64\,$a.u. for the longest pulse, whereas both minima occur for intermediate pulse durations. Note that $\tau$ refers to 
the FWHM of the intensity profile.
The results displayed in Fig.~\ref{fig:plot4} (top) already provide an insight into possible mechanisms for enhancing the degree of coherence between the $3s$ and $3p_0$ hole states: For perfect hole coherence, photoelectrons from the $3s$ and $3p_0$ orbitals must be energetically indistinguishable. 
The binding energy is $1.272\,$a.u. for  $3s$ and $0.591\,$a.u. for
$3p_0$
at the Hartree-Fock level. Therefore, a photon with $\omega_{ph}=0.50\,$a.u. might  create, via three-photon ionization of the $3s$ orbital, a photoelectron at an energy of $\omega_{e^-(3s)}=0.228\,$a.u. while 
two-photon ionization of the $3p_0$ orbital would create a photoelectron at  $\omega_{e^-(3p_0)}=0.409\,$a.u. This is one scenario, where the minimum bandwidth required for energetic indistinguishability corresponds to a maximum $\tau=30.7\,$a.u.
This scenario corresponds to the minimum in Fig.~\ref{fig:plot4} (top) near $\omega_{ph}=0.50\,$a.u. for $\tau$ up to 35$\,$a.u. For shorter pulses, the minimum becomes broader but  remains centered at $\omega_{ph}=0.50\,$a.u.
The second minimum, near $\omega_{ph} = 0.64\,$a.u., observed for long and  spectrally narrow pulses, cannot be explained by this first scenario. For example,  $\tau=47\,$a.u. corresponds to a spectral bandwidth of $0.06\,$a.u. 
However, a central frequency of $\omega_{ph} = 0.64\,$a.u. is not too far from the transition frequency between the parent orbitals,  $\delta\omega_{3s,3p_0}= 0.681\,$a.u. A second conceivable scenario thus consists in the one-photon ionization of the $3p_0$ orbital together with the resonant excitation of a $3s$ electron into the $3p_0$ hole. 
One-photon ionization of the $3p_0$ orbital with a photon of $\omega_{ph}=0.64\,$a.u. would lead to a photoelectron at $E_{e^-(3p)} = 0.049\,$a.u., whereas a photoelectron originating from the $3s$ orbital that absorbed two such photons would have an energy of $E_{e^-(3s)}=0.008\,$a.u.

\begin{figure}[tb]
\centering
\includegraphics[width=0.95\linewidth]{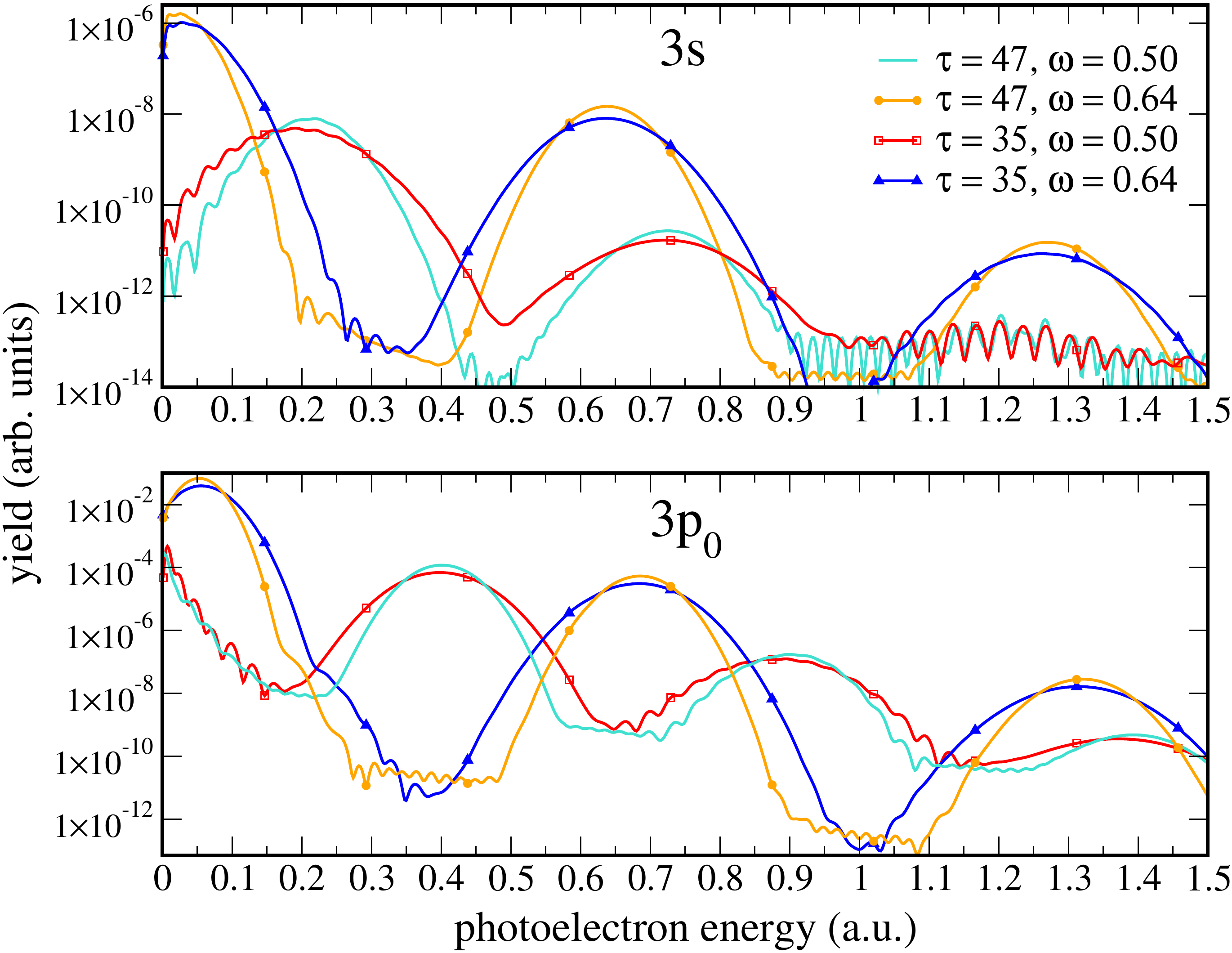}
\caption{Channel-resolved PES obtained from the transform-limited Gaussian pulses studied in Fig.~\ref{fig:plot4} for a maximal field amplitude of $E_0 = 0.02\,$a.u.}  
\label{fig:plot4_channel_pes}
\end{figure}
In order to check whether these scenarios are indeed responsible for the structure observed in Fig.~\ref{fig:plot4} (top),    
channel-resolved photoelectron spectra (PES) are shown in Fig.~\ref{fig:plot4_channel_pes}. Indeed, for $\omega_{ph}=0.64\,$a.u. and $\tau = 47\,$a.u. (yellow lines), the channel-resolved PES reveal for $3s$ a peak in the vicinity of $\omega_{e^-(3s)} = 0.01$ and for $3p_0$ one at $\omega_{e^-(3p_0)} = 0.05\,$a.u. Given our resolution, these peaks essentially coincide with the expected ones at 0.008$\,$a.u. and $0.049\,$a.u., confirming the creation of hole coherence by resonant transition from the $3s$ into the $3p_0$ orbital. The same mechanism is seen to be at work for the pulse with $\tau=35\,$a.u. and $\omega_{ph}=0.64\,$a.u. (dark blue line in Fig.~\ref{fig:plot4_channel_pes}). The larger width of the blue peaks compared to the yellow ones ($\tau = 47\,$a.u.) simply reflects the larger bandwidth of the field. 

Completely different PES are obtained for a central frequency of  $\omega_{ph} = 0.50\,$a.u. (red and cyan lines in Fig.~\ref{fig:plot4_channel_pes}). Assuming here the first scenario to be relevant, i.e., a simultaneous three-photon ionization of $3s$ and two-photon ionization of $3p_0$, we expect peaks at
$\omega_{e^⁻(3s)}=0.228\,$a.u. in the $3s$-PES and at $\omega_{e^-(3p)}=0.409\,$a.u. in the $3p_0$-PES. These peaks are indeed observed for the red and cyan curves in Fig.~\ref{fig:plot4_channel_pes}. Even if for  $\tau=47\,$a.u. (cyan line in Fig.~\ref{fig:plot4_channel_pes}) the spectral bandwidth is too small to really render the $3s$ and $3p_0$ photoelectrons indistinguishable, 
the mechanism of simultaneous three-photon ionization of $3s$ and two-photon ionization of $3p_0$ explains the small dip at $\omega_{ph}=0.50\,$a.u. in the brown line in Fig.~\ref{fig:plot4}. This  holds of course also for the deeper minima observed for shorter, i.e., spectrally broader pulses. 
We thus conclude that the first scenario, of simultaneous three-photon ionization of $3s$ and two-photon ionization of $3p_0$, is at work for $\omega_{ph} = 0.50\,$a.u.

\begin{figure}[tb]
\centering
\includegraphics[width=0.95\linewidth]{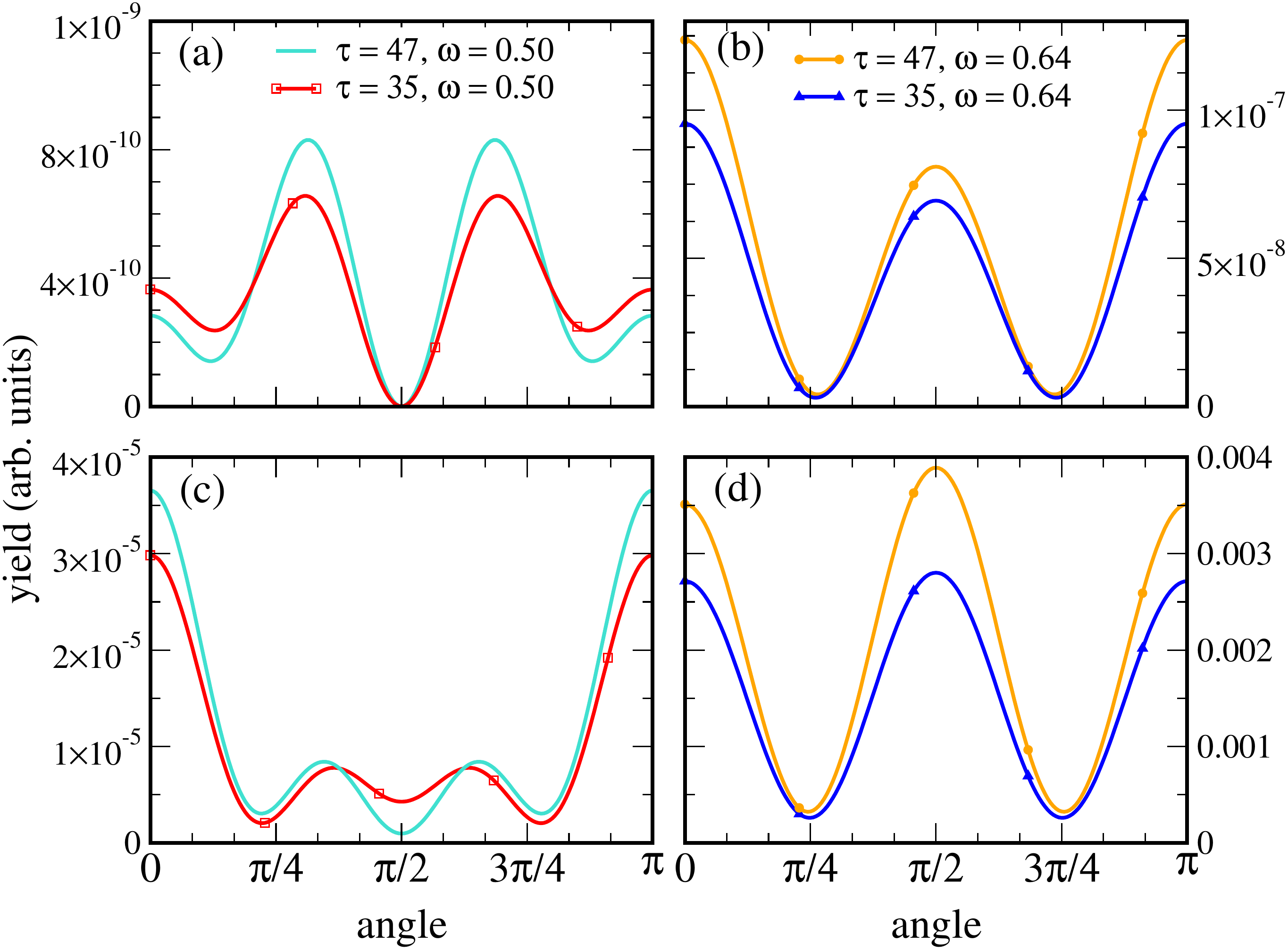}
\caption{Channel-resolved PAD corresponding to the PES shown in Fig.~\ref{fig:plot4_channel_pes}: Panels (a) and (b) display the
contribution of $3s$ photoelectrons to the energy-integrated PAD, panels (c) and (d) that of $3p_0$ photoelectrons.}  
\label{fig:plot4_channel_pad}
\end{figure}
For completeness, the channel-resolved energy-integrated photoelectron angular distributions (PADs) corresponding to the PES of Fig.~\ref{fig:plot4_channel_pes} are shown in Fig.~\ref{fig:plot4_channel_pad}. Interestingly, for the first control scenario, the angular distributions are completely different for $3s$ and $3p_0$ photoelectrons, whereas they are very similar for the second one. 
This is not too surprising since in the second control scenario, 
the $3s$ electron is, after creation of a $3p_0$ hole, resonantly
excited into the
$3p_0$ orbital before being ionized. In contrast, in the first control
scenario,  $3s$ and $3p_0$ electrons are directly ionized which renders a correlation between the $3s$ and $3p_0$ PADs more unlikely.

A scan of the central frequency thus provides not only a good initial value for this parameter but also insight into the possible control mechanisms. 
A more complete picture is obtained when scanning both frequency and
peak amplitude of the field, keeping only the duration fixed. The
results are shown in Figs.~\ref{fig:plot4}(b) and (c)
for pulse durations of $\tau=6\,$a.u. and  $\tau=23\,$a.u., respectively: Apparently, spectrally too broad pulses are not suitable for the maximization of hole coherence, cf.~Fig.~\ref{fig:plot4}(b).
The best pulses are obtained for  $\tau=23\,$a.u. (light-blue area in Fig.~\ref{fig:plot4}(c)) where a distinct window  of favorable central circular frequencies occurs between $\omega_{ph} = 0.50\,$a.u. and $\omega_{ph} = 0.65\,$a.u. Interestingly, good hole coherences are obtained even for weak fields. One has to keep in mind, however, that these come with low overall ionization probabilities. 

\begin{figure}[tb]
\centering
\includegraphics[width=0.95\linewidth]{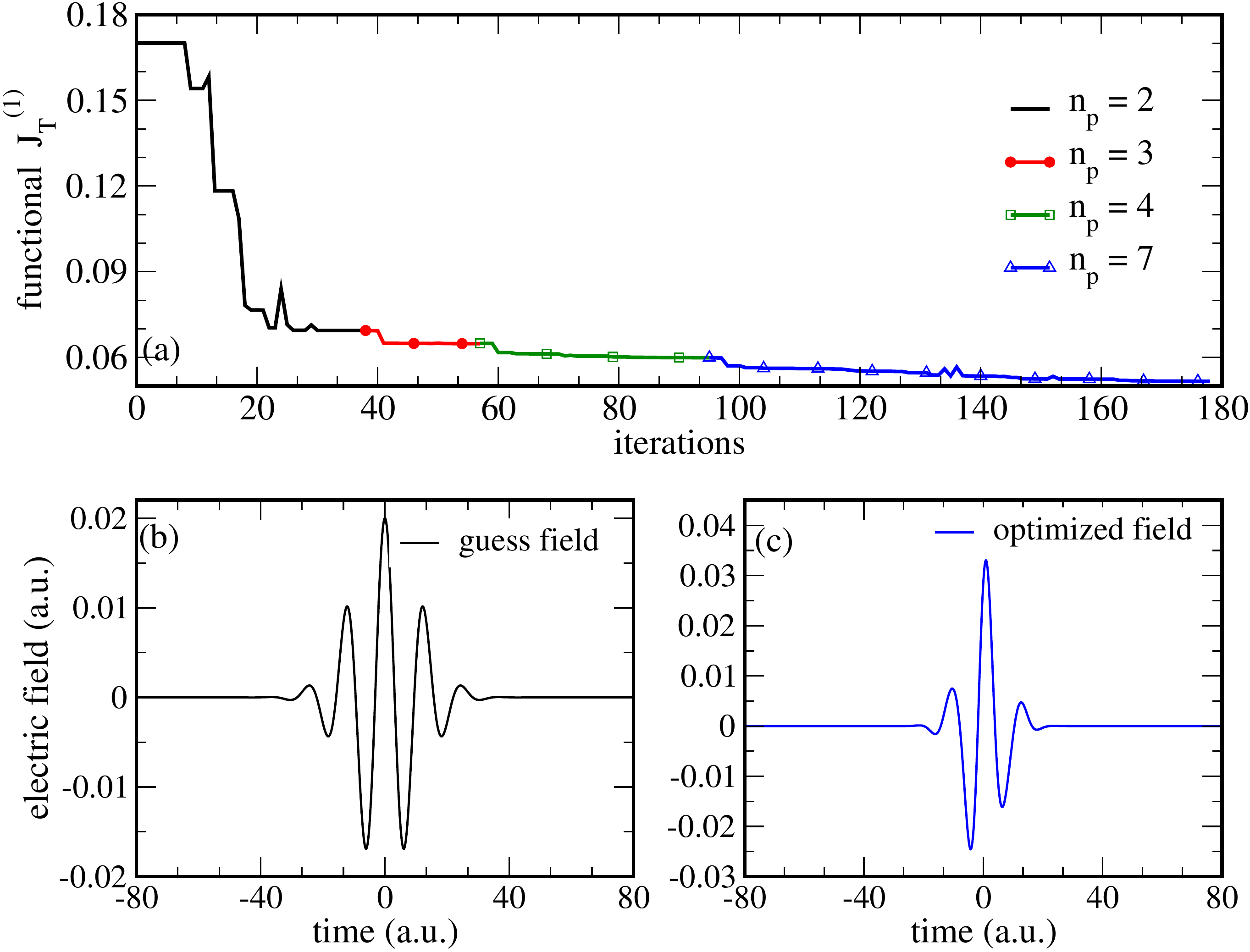}
\caption{
  SPA-optimization with the principal axis method, using favorable
  initial parameters in the guess pulse: The convergence is
  significantly accelerated (a). Guess and optimized fields are
  shown in (b) and (c).
}  
\label{fig:plot5}
\end{figure}
Once we have scanned the basic parameters of the field, we use the best values to start the actual SPA-optimization, increasing the number of parameters once the change in the 
functional, $J_T^{(1)}$, becomes too small, as before. Figure~\ref{fig:plot5} shows the 
corresponding results. The parameter scan allows to find an already good guess field, depicted in Fig.~\ref{fig:plot5}(b), such that SPA-optimization starts with a value of $J_T^{(1)}=0.17$, cf. Fig.~\ref{fig:plot5}(a), to be compared with the poor starting fidelity in Figs.~\ref{fig:plot1a} and~\ref{fig:plot2}.
After only 180 iterations, $J_T^{(1)}$ has dropped to $0.04$.
At this stage, 7 optimization parameters are used, resulting in a comparatively simple shape of the optimized field, cf. Fig.~\ref{fig:plot5}(c). For comparison, the lowest value of $J_T^{(1)}$ obtained in Sec.~\ref{subsec:arbitrary} without a prior parameter scan amounts to 
$0.07$. 
Thus, the sequential update technique based on the principal axis method, with prior scanning of the optimal parameters for the guess field reveals itself to be a very efficient
optimization method. It allows for reaching high fidelities while minimizing
the number of optimization parameters as well as the numerical effort. 

\begin{figure}[tb]
\centering
\includegraphics[width=0.95\linewidth]{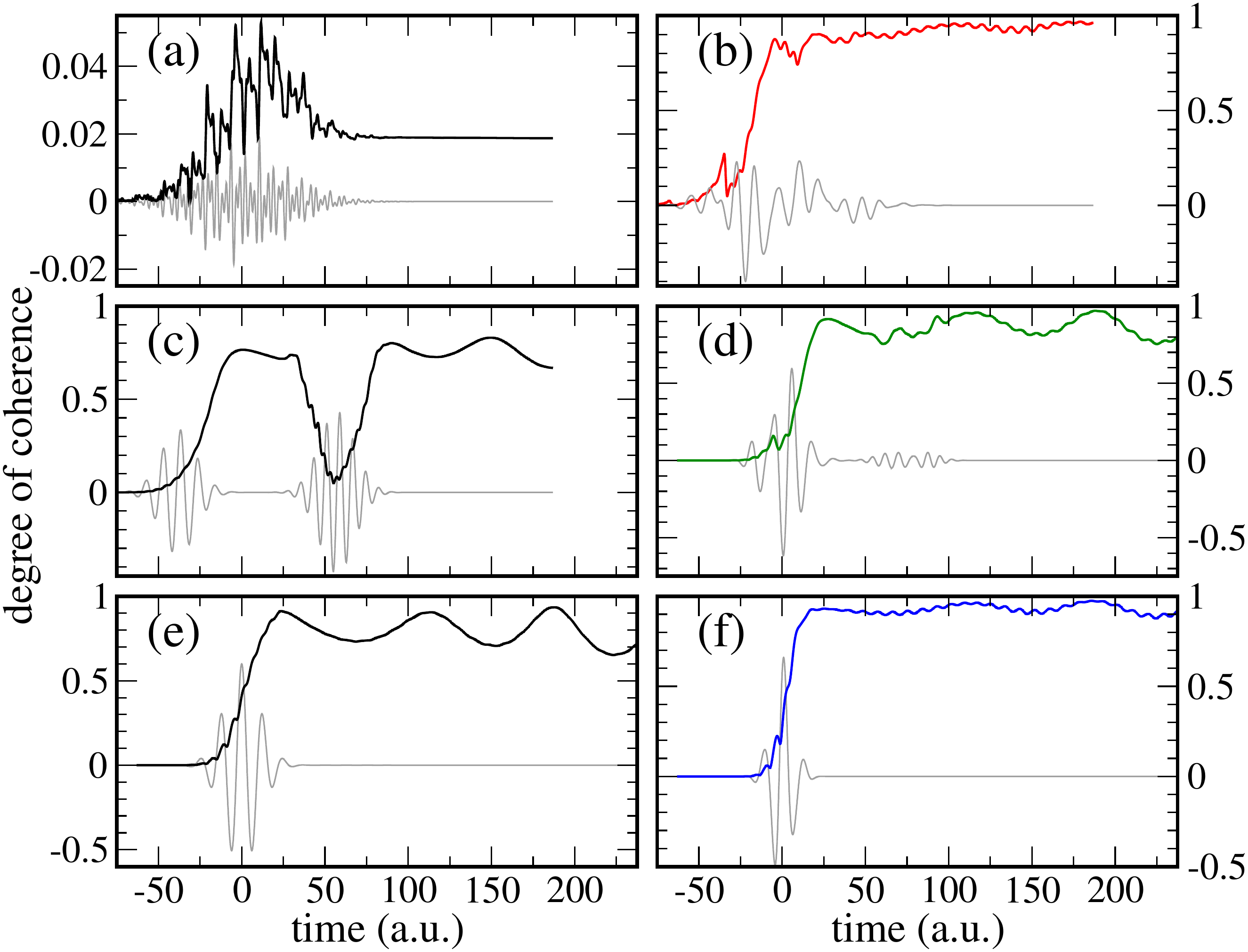}
  \caption{Degree of coherence as a function of time obtained 
    with guess (left) and optimized (right) fields: (a) randomly
    chosen initial parameters ($n_p=26$); (b) corresponding optimized
    field with $g_{3s,3p_0}(T)=0.989$ ($n_p=26)$; 
    (c) initial guess field consisting of two time-delayed Gaussians
    ($n_p=8$) 
    (d) corresponding optimized
    field  for which the degree of coherence
    oscillates between $g_{3s,3p_0}(t)=0.97$
    and $0.75$ with a final value of $g_{3s,3p_0}(T)=0.80$ ($n_p=16)$; 
    (e) initial monochromatic guess field with favorable parameters
    identified by parameter scan, 
    (f) corresponding optimized
    field ($n_p=7$, the same as shown in
    Fig.~\ref{fig:plot5}(c)) for which the degree of coherence
    oscillates between $g_{3s,3p_0}(t)=0.98$
    and $g_{3s,3p_0}(T)=0.90$.             
}
\label{fig:plot6}
\end{figure}
The dynamics obtained with various guess and optimized fields are analyzed in 
Fig.~\ref{fig:plot6}, which displays the degree of coherence as a function of time. Figures~\ref{fig:plot6}(a) and (b) compare $g_{3s,3p_0}(t)$
for a randomly chosen guess field with a large number of parameters (black line) and for the optimized field  obtained from this guess (red line). The fields are shown in grey (not scaled). Whereas the guess field yields a very poor fidelity, cf. the y-axis scale, the maximized degree of coherence between the hole states $3s$ and $3p_0$, reaches a value of $g_{3s,3p_0}=0.989$.
Figures~\ref{fig:plot6}(c) and (d) answer the question whether a time-delayed sequence of two Gaussian pulses is suitable for maximizing hole coherence. We treat the amplitudes, circular frequencies and delay as a optimization parameters.
Since the subpulse structure essentially disappears upon optimization, we conclude that time-delayed pulses are not suitable for maximizing hole coherence.
Finally, Figs.~\ref{fig:plot6}(e) and~\ref{fig:plot6}(f) display the degree of coherence obtained with the guess constructed after parameter scan and the corresponding optimized field, also shown in Fig.~\ref{fig:plot5}(b) and (c).

\begin{figure}[tb]
\centering
\includegraphics[width=0.95\linewidth]{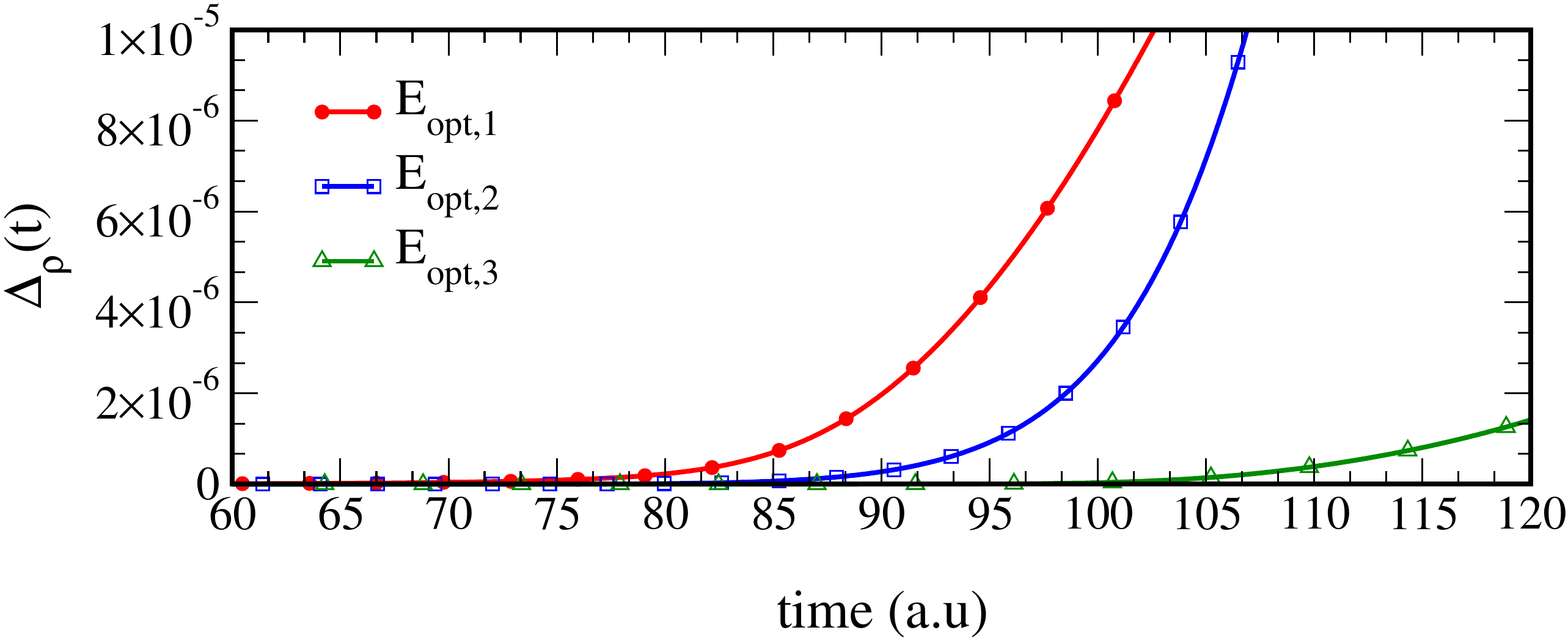}
\caption{Photoionization probability, obtained in terms of the
  absorbed part of  the ion density matrix, cf.~Eq.~\eqref{eq:Delta}, 
  as a  function of time for the three optimized fields
    depicted in Fig.~\ref{fig:plot6}(b), (d) and (f). The color code
    is the same as in Fig.~\ref{fig:plot6}.
}
\label{fig:plot7}
\end{figure}
Remarkably, the degree of coherence oscillates as a function of time in Fig.~\ref{fig:plot6}, even after the 
field is over. These oscillations may be related to
  two possible mechanisms: On one hand, the oscillations might be
  related to how fast the photoelectron leaves the parent ion
  since the interaction between 
  any outgoing photoelectron and the remaining ion creates
  entanglement and thus decreases the hole coherence.  On the other
  hand, they may be caused by excitation of Rydberg states, which
  would allow the electron-ion interaction to persist even long after
  the pulse is over.  In both cases, the excited electron 
  reaches a sufficiently large spatial extension to be affected by the
                     CAP.
To analyze how fast the excited electron reaches the
region of the CAP, Fig.~\ref{fig:plot7} shows the 
correction to the ion density matrix due to the CAP, cf. Eq.~\eqref{eq:Delta}, for the three different optimized fields shown on the right-hand side of Fig.~\ref{fig:plot6}. The optimized field, for which the degree of coherence shows the fastest oscillations 
with the smallest amplitude (red line in  Fig.~\ref{fig:plot6}), produces the more energetically excited electrons (the ones reaching large spatial domain first), whereas
the slowest oscillations of the degree of coherence with the largest amplitude (green line in Fig.~\ref{fig:plot6}) are associated with the 
less energetically excited electrons reaching the CAP region, cf.
Fig.~\ref{fig:plot7}. From these observations we may conclude that the oscillations arise from the interaction between the remaining ion and the excited electron, which perturbs the coherence of the ion density matrix. 
Thus, the fastest excited electrons interact the
least with the remaining ion whereas the slowest 
(or bound) ones, which interact with the remaining ion during longer times, lead to a larger perturbation of 
the degree of coherence. A similar conclusion regarding the interaction between the photoelectron and the photoion was previously drawn for hole decoherence in the photoionization of xenon~\cite{PabstPRL2011}.

\begin{figure}[tb]
\centering
\includegraphics[width=0.95\linewidth]{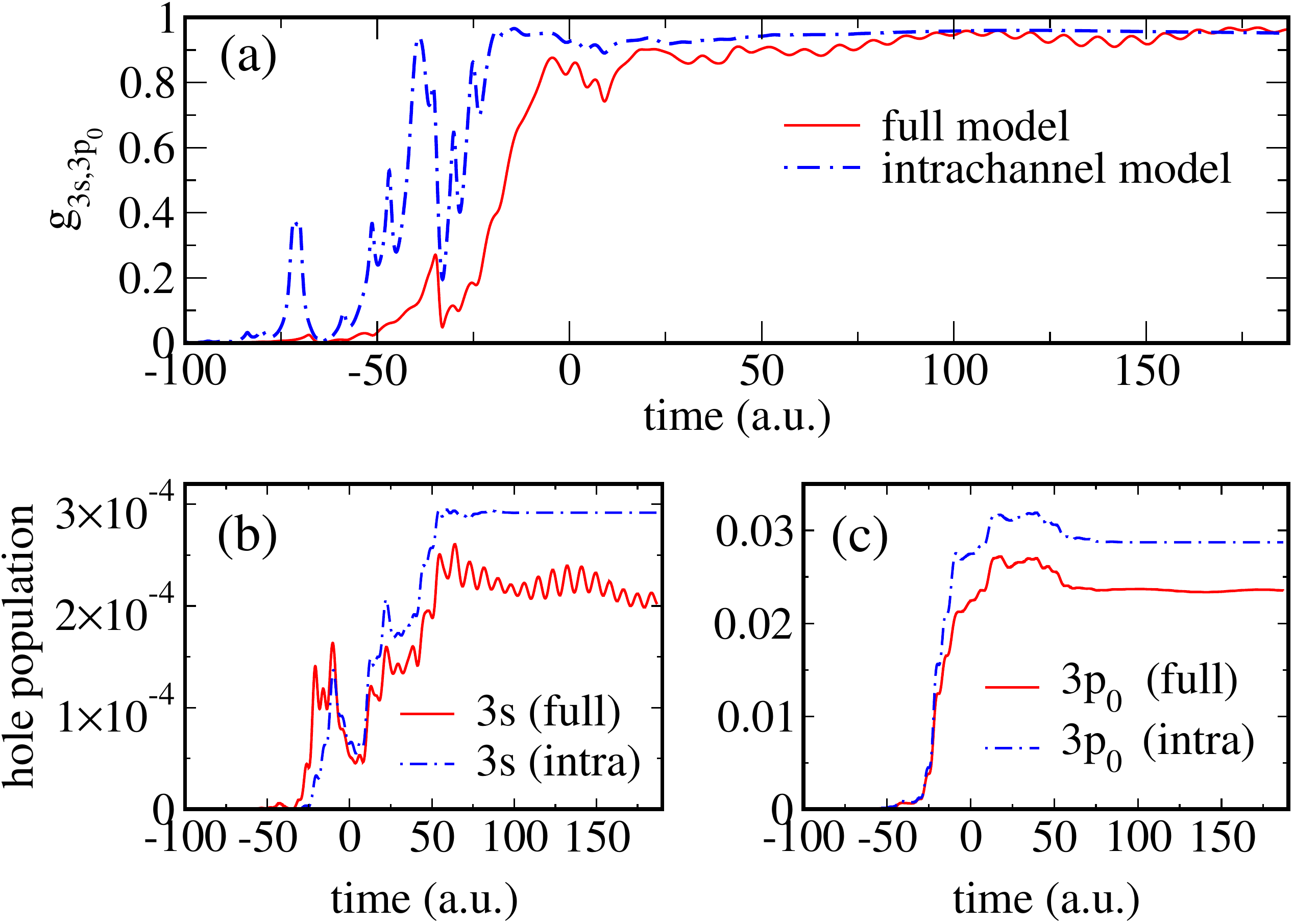}
\caption{Degree of coherence (a) and hole populations (b,c) as a
  function of time, 
  obtained with the optimized field shown in Fig.~\ref{fig:plot6}(b), 
  for the interchannel ('full')  and intrachannel models.}
\label{fig:plot8.5_b}
\end{figure}
\begin{figure}[tb]
\centering
\includegraphics[width=0.95\linewidth]{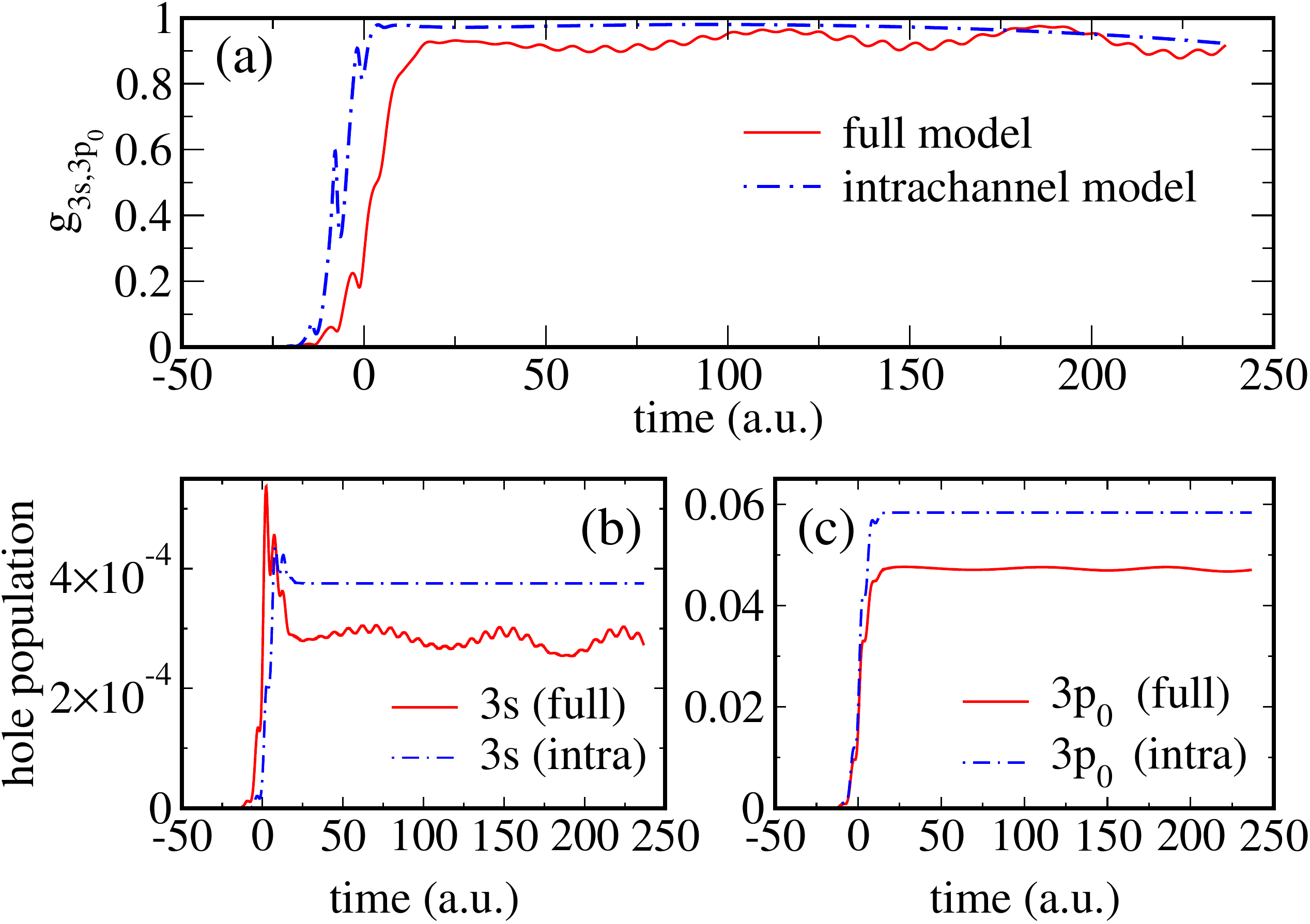}
\caption{Degree of coherence (a) and hole populations (b,c) as a
  function of time, 
  obtained with the optimized field shown in Fig.~\ref{fig:plot6}(f), 
  for the interchannel ('full')  and intrachannel models.}
\label{fig:plot8.5_f}
\end{figure}
This interpretation is relevant for the ``full'' model including interchannel coupling where a fast departure of the photoelectron minimizes 
the interaction with the remaining ion. In contrast, within the intrachannel model, the excited electron can interact only with the  electrons remaining in the channel from which it originates. One should therefore expect that the oscillations in this  case become less important. In Fig.~\ref{fig:plot8.5_b}, we compare the degree of coherence as well as the hole populations as a function of time for the ``full'' model and the 
intrachannel approximation. We have used the optimized field, depicted in
Fig.~\ref{fig:plot6}(b), that produces the fastest photoelectrons (within the ``full'' model), so that the oscillations in $g_{3s,3p_0}$ are minimal. As can be seen in Fig.~\ref{fig:plot8.5_b}(a), the
oscillations at times larger than 50$\,$a.u., due to the interaction between the excited electron and the parent ion, dissapear completely if we allow the excited electron to interact only with the orbital from which it originates.
Furthermore, the interchannel coupling  is also found to be responsible for the oscillations in the hole populations after the pulse is over, cf.
Fig.~\ref{fig:plot8.5_b}(b) and  Fig.~\ref{fig:plot8.5_b}(c). 
In Fig.~\ref{fig:plot8.5_f}, we carry out the same analysis of the interchannel coupling, this time using the optimized field depicted in Fig.~\ref{fig:plot6}(f), which produces slower photoelectrons, cf. Fig.~\ref{fig:plot7}. Again, the oscillations in $g_{3s,3p_0}(t)$ 
and the hole populations, observed for the ``full'' model, disappear in the intrachannel approximation. Despite the modified dynamics, the final value for the degree of coherence remains almost the same  for both optimized fields when switching off the interchannel coupling. In contrast, the final hole populations are considerably changed, cf. the lower panels in Figs.~\ref{fig:plot8.5_b} and \ref{fig:plot8.5_f}.
This strongly suggests that the oscillations present in the degree of coherence as well as in the hole populations are induced by the interchannel interaction.

\section{Maximization of the coherence with prescribed hole population target}
\label{sec:prescribed_population}

A remarkable feature of the optimization results presented in the previous section is the population difference between the hole states. Indeed, 
the population of the $3p_0$ hole exceeds that of the $3s$ hole by at least two orders of magnitude in all examples studied. Such a large population difference is undesirable in view of utilizing the coherent superposition in time-dependent spectroscopy. We therefore address now the question whether it is possible to maximize the degree of coherence between the $3s$ and $3p_0$ hole states while simultaneously controlling the final hole population. 

\begin{figure}[tb]
\centering
\includegraphics[width=0.95\linewidth]{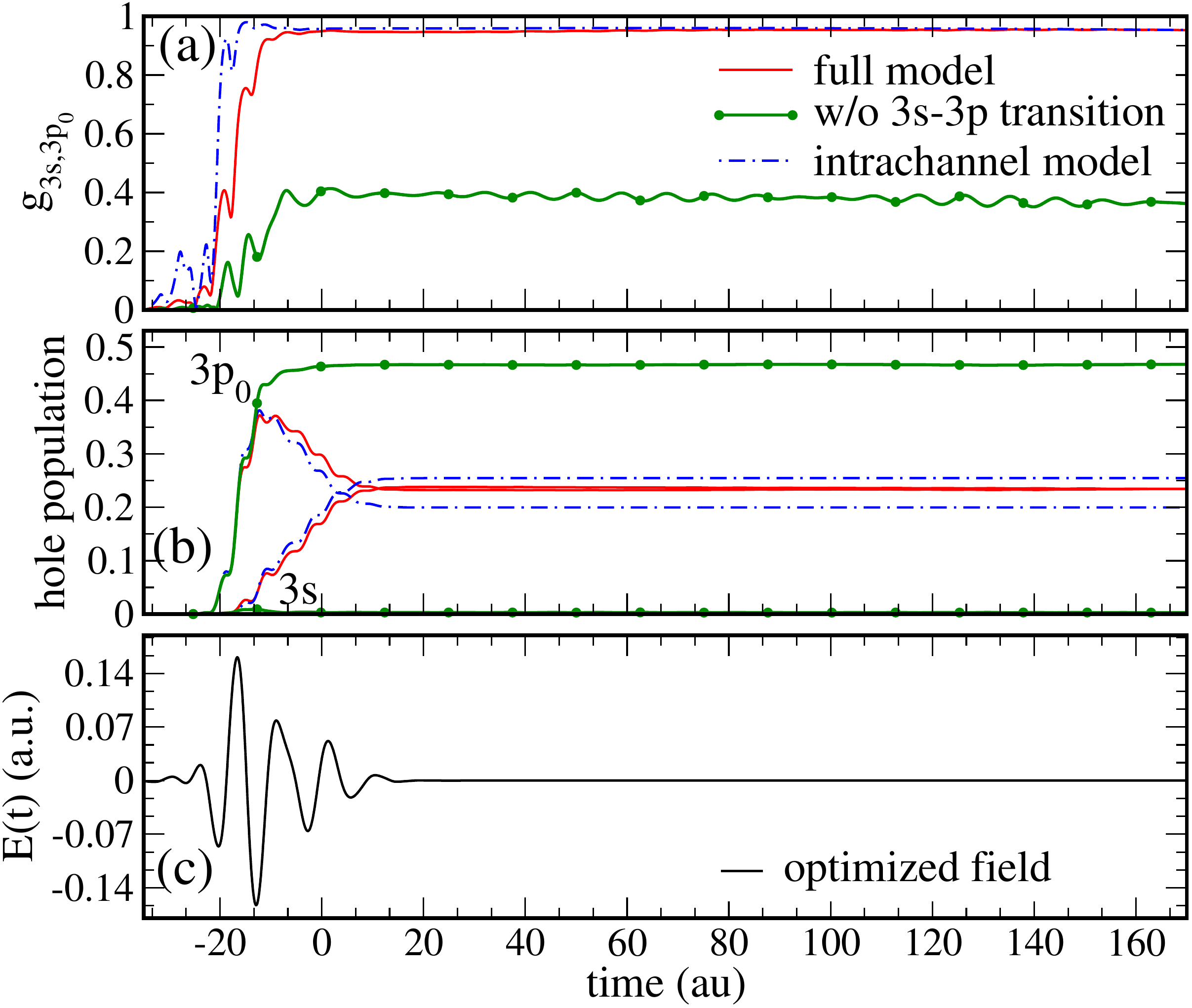}
  \caption{Maximizing the degree of coherence between the $3p_0$ and
    $3s$ hole states while simultaneously optimizing for a hole 
    population ratio of one: degree of coherence (a),
    hole populations (b) and optimized electric field (c) as a
    function of time. 
}
\label{fig:application_coherence_t:02}
\end{figure}
We consider all possible scenarios, i.e., equal populations, $\rho_{3p_0,3p_0} > \rho_{3s,3s}$, and $\rho_{3p_0,3p_0} < \rho_{3s,3s}$. To be specific, we ask for the corresponding population ratio $\mathcal R$ to be equal to 0.7 in the last two cases and utilize the optimization functional $J^{(2)}_T$, cf.~Eq.~\eqref{eq:functionalf}. 
Starting with equal populations, Fig.~\ref{fig:application_coherence_t:02} shows the degree of coherence, hole populations and optimized field as a function of time, demonstrating success of SPA-optimization also for this more challenging control target. Figure~\ref{fig:application_coherence_t:02} also analyzes the role of the interchannel coupling, cf. red and blue lines, as well as the role of direct transitions between the $3s$ and $3p_0$ states, cf. red and green lines. The interchannel coupling is seen to affect the hole coherence only during the first half of the pulse, whereas the final coherence is identical with and without interchannel coupling, cf. Fig.~\ref{fig:application_coherence_t:02}(a). In contrast, suppressing the excitation of a $3s$ electron into the $3p_0$ orbital strongly modifies the degree of coherence. It reduces the final value from 0.98 to 0.39, 
indicating that sequential ionization of $3s$ electrons is important here. 

As for the population dynamics, Fig.~\ref{fig:application_coherence_t:02}(b) reveals the $3p_0$ hole population to always be larger than the $3s$ population until the two populations reach the same value. This is true both with and without interchannel coupling. The interchannel coupling is seen to only affect the final populations, by an amount that is not very large.
While the $3s$ hole population increases
monotonically, the $3p_0$ hole population reaches a maximum value at the same time that the degree of coherence becomes stationary. After that time, the $3p_0$ hole population decreases to the target value. In contrast to the degree of coherence that becomes stationary already while the pulse is still on, the hole populations do so only at the end of the pulse. The population dynamics confirms the importance of excitations from $3s$ electrons to $3p_0$: When this transition is switched off, the $3s$ hole population drops to essentially zero, cf. the green line in Fig.~\ref{fig:application_coherence_t:02}(b). We can thus conclude that the decrease of the  $3p_0$ hole population and simultaneous increase of the $3s$ hole population, seen for the ``full'' model, is due to a dipole transition between these 
two states. In other words, Rabi oscillations occur between these orbitals, as indicated by the oscillatory pattern of the red and blue lines in Fig.~\ref{fig:application_coherence_t:02}(b) for $-10\le t\le 15\,$a.u. 
This interpretation is confirmed by the fact that these oscillations occur with the same frequency, but a phase shift of $\pi$ (data not shown).
 
\begin{figure}[tb]
\centering
\includegraphics[width=0.95\linewidth]{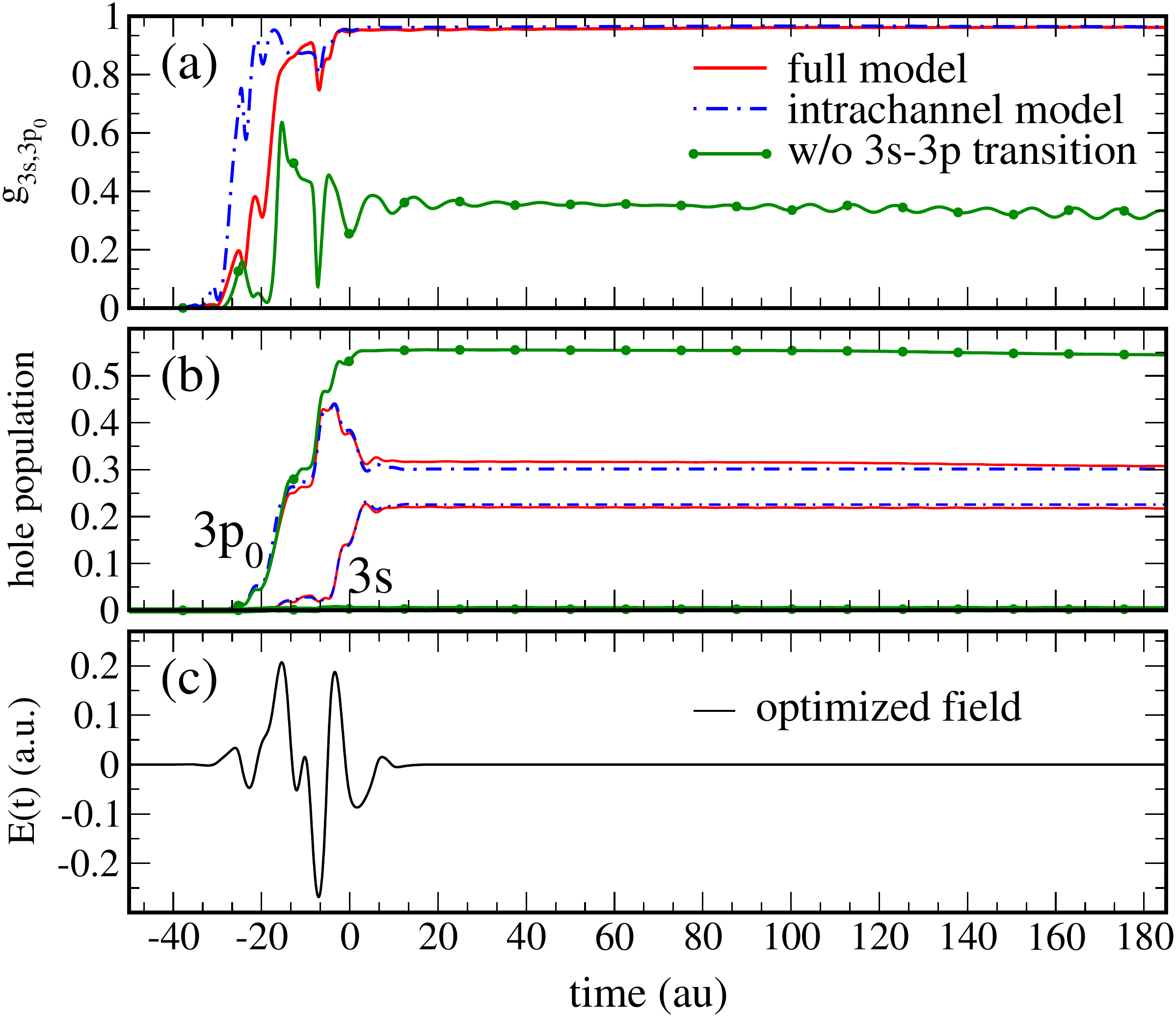}
  \caption{
    Maximizing the degree of coherence between the $3p_0$ and
    $3s$ orbitals while simultaneously optimizing for a hole 
    population ratio of $\rho_{3s,3s}/\rho_{3p_0,3p_0}=0.7$:
    degree of coherence (a),
    hole populations (b) and optimized electric field (c) as a
    function of time. 
}
\label{fig:application_coherence_t:01}
\end{figure}
Next, we target the case $\rho_{3p_0,3p_0} > \rho_{3s,3s}$ with a population ratio of  $\mathcal R=0.7$. Given the fact that the $3p_0$ hole population always turned out to be larger than the $3s$ hole one in Sec.~\ref{sec:optimization_results}, this is the simplest of the three cases. The results are shown in 
Fig.~\ref{fig:application_coherence_t:01}.   
Similarly to the case of equal hole populations, the interchannel coupling does not affect the final degree of coherence and the final populations. However, in contrast to the case of equal populations, both the hole
population and the degree of coherence become stationary at the same
time, once the pulse is over, cf. Fig.~\ref{fig:application_coherence_t:01}(a) and~(b). Direct transitions between the $3s$ and $3p_0$ orbitals are found to play again an important role, cf. the green lines in Fig.~\ref{fig:application_coherence_t:01}(a) and~(b).

\begin{figure}[tb]
\centering
\includegraphics[width=0.95\linewidth]{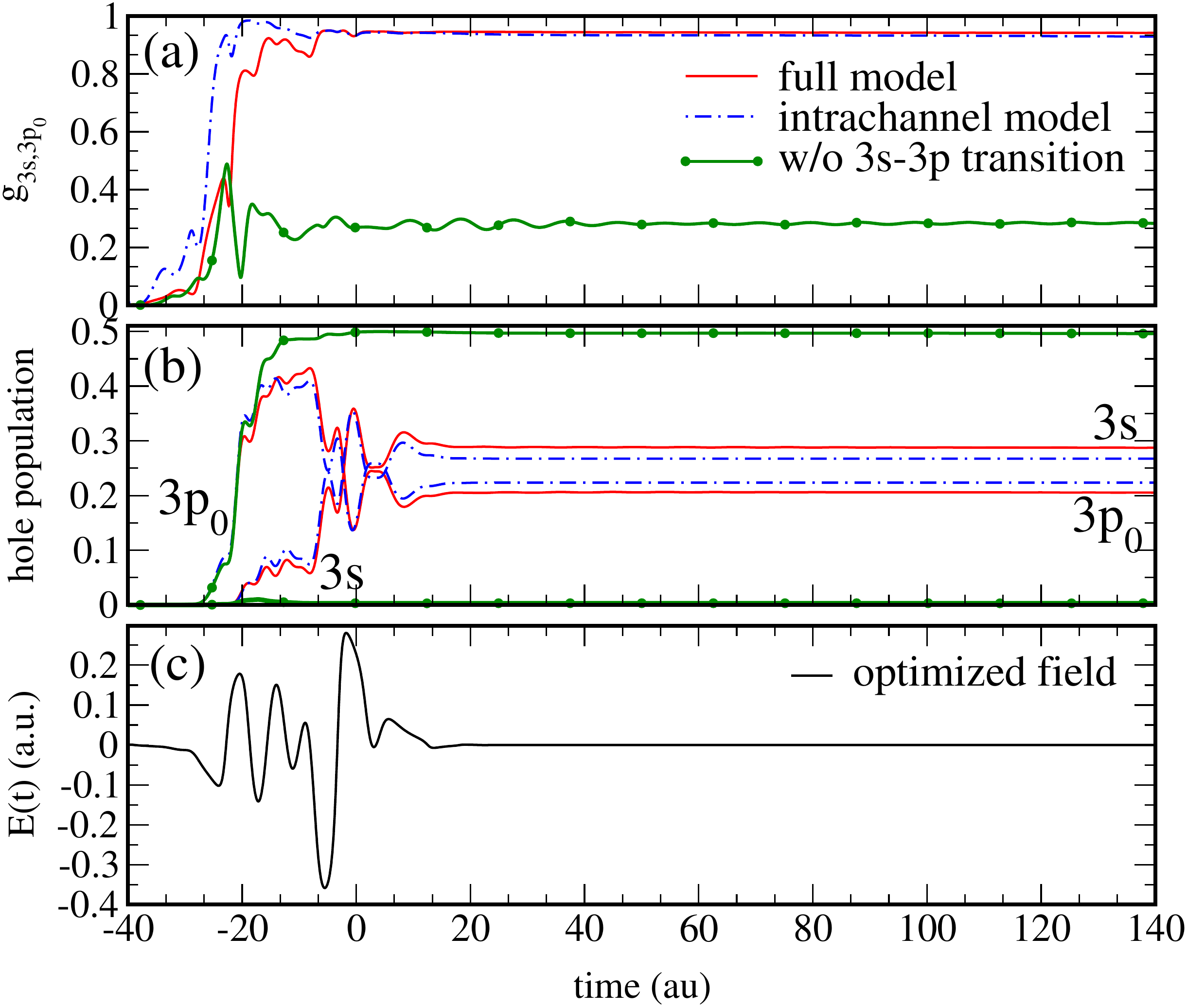}
  \caption{
   Maximizing the degree of coherence between the $3p_0$ and
   $3s$ hole states while simultaneously optimizing for a hole 
    population ratio of $\rho_{3p_0,3p_0}/\rho_{3s,3s}=0.7$:
    degree of coherence (a),
    hole populations (b) and optimized electric field (c) as a
    function of time. 
}
\label{fig:application_coherence_t:03}
\end{figure}
Finally, we maximize the degree of coherence constraining the hole populations such that $\rho_{3p_0,3p_0} < \rho_{3s,3s}$. This is the most difficult target, but it is successfully addressed by SPA-optimization and the results are shown in Fig.~\ref{fig:application_coherence_t:03}.
Again, the interchannel coupling is found to affect the degree of coherence only during the pulse, but neither the final coherence nor the population dynamics, cf. red and blue lines in Fig.~\ref{fig:application_coherence_t:03}(a) and~(b).
Compared to the cases of equal population and larger $3p_0$ hole population, the population dynamics is more intricate, showing a crossing in order to reach the desired population ratio and a number of distinct oscillations. 
We again check whether these oscillations correspond to Rabi cycling 
between the $3s$ and $3p_0$ orbitals by switching off the transition dipole matrix elements. We find that, when $3s$ to $3p_0$ transition are not allowed, no oscillations are present in the population dynamics, and the $3s$ hole population drops to essentially zero. Moreover, analysis of the population oscillations reveals again their identical frequency and a phase shift of $\pi$ (data not shown). 

\begin{figure}[tb]
\centering
\includegraphics[width=0.95\linewidth]{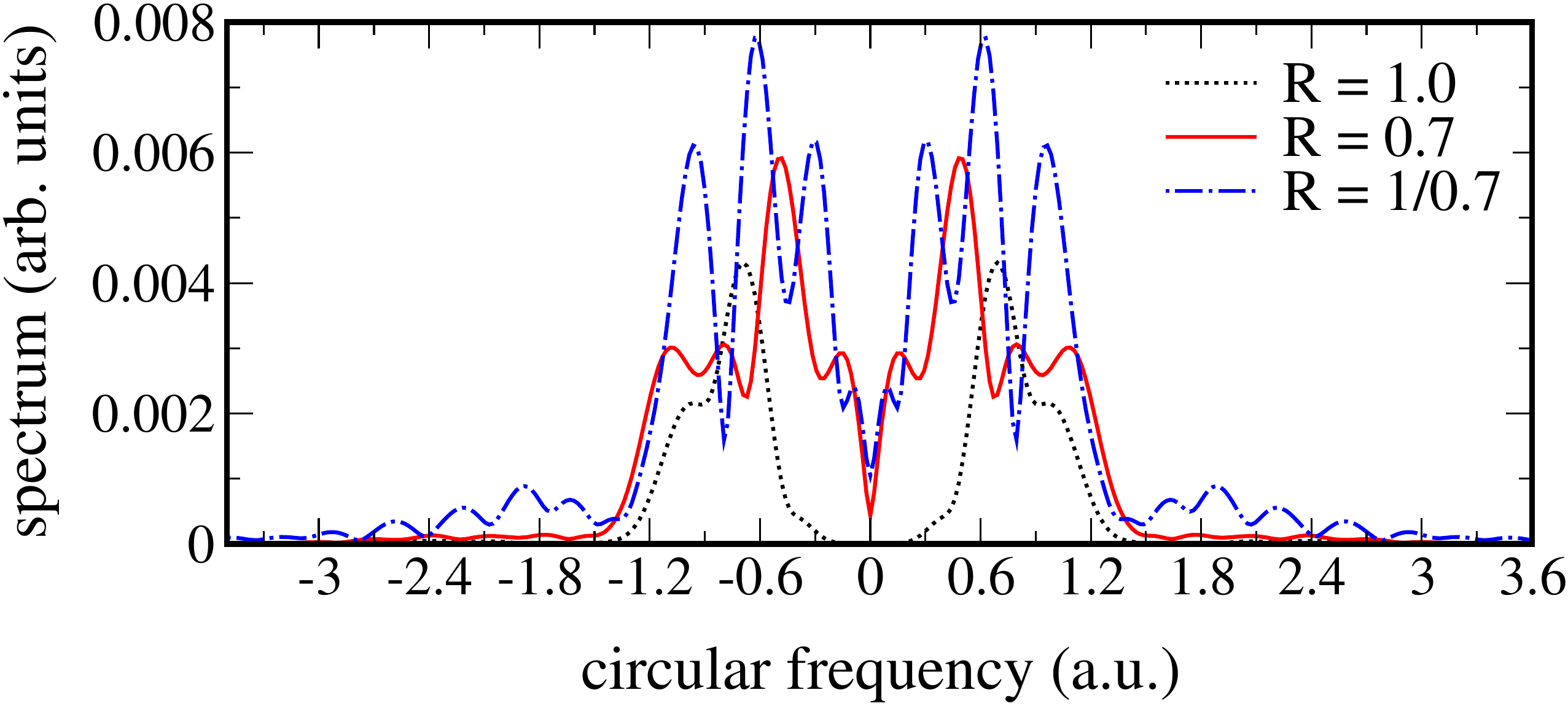}
  \caption{
    Maximizing the degree of coherence between the $3p_0$ and
    $3s$ hole states while simultaneously optimizing for a given hole 
    population ratio: Spectra of the optimized fields for the three different 
    hole population ratios, $\mathcal{R} = \rho_{3s,3s}/\rho_{3p_0,3p_0}$.
}
\label{fig:optimized_fields_spectrum}
\end{figure}
For all three variants of the $3s$ to $3p_0$ hole population ratio, the corresponding optimized fields were successfully identified by SPA-optimization. Their 
spectra are shown in Fig.~\ref{fig:optimized_fields_spectrum}. The
circular frequencies  were treated as 
optimization parameters, using Eq.~\eqref{eq:freq_map} to constrain them to
$\omega_{min} =-4\,$a.u. and $\omega_{max} =4.0\,$a.u.
The most difficult optimization target results in the broadest spectrum, cf. blue line in Fig.~\ref{fig:optimized_fields_spectrum}. It is a common observation that more difficult optimization problems result in more complex control fields.
Overall, the optimized spectra are too broad to identify one of the two control mechanisms, based on photon energies of 0.50$\,$a.u. versus 0.68$\,$a.u., as discussed in the previous section, by inspection of the spectra alone. 
The numerical effort, in terms of optimization parameters is comparable for all three cases---the final number of optimization parameters amounts to $28$.
The most difficult optimization target required the largest number of iterations. In this case, the value of the functional $J^{(2)}_T$
decreased with a slower rate, compared to the other two cases.  
For all three population ratios, SPA-optimization was started with the same guess field, using four optimization parameters: the FWHM, a frequency, a Fourier amplitude and a phase shift. At the end of the procedure, the FWHM, nine frequency components, nine Fourier amplitudes and nine phases were optimized.

\section{Conclusions}
\label{sec:concl}

To summarize, we have introduced a sequential update of the pulse parametrization to ease implementation of gradient-free parameter optimization in quantum control. We have applied this technique to maximize the coherence of hole state superpositions in the photoionization of argon. A sequential update of the pulse parametrization, which adds more terms to the parametrization once the optimization gets stuck, allows for faster convergence and better final results. Such a sequential update can be combined with any method for parameter optimization, and we have tested it here for the principal axis method and the Nelder Mead downhill simplex approach. The principal axis method which so far has not been employed in quantum control turns out to be clearly more efficient than the
widely used Nelder Mead approach. Thus, the principal axis method, in particular when combined with a sequential parametrization update, 
represents an efficient and viable tool for quantum control.

Admittedly, parameter optimization comes with the disadvantage of depending, sometimes critically, on the chosen parametrization. This is outweighted in our case by the ease of implementation, even for a non-Hermitian Hamiltonian.  The latter is due to the fact that the long propagation times for photoionization require the use of a complex absorbing potential. For comparison, the alternative approach of gradient-based optimization always involves backward-in-time propagation of Lagrange multiplier wavefunctions, and the CAP becomes, in the adjoint equation, a source term which can easily give rise to numerical instability. 

The technique introduced here can be further improved by scanning key parameters prior to optimization. The numerical effort required for the scan is more than paid off by the reduction in the number of iterations. It also allows for an identification of possible control mechanisms. In our example, determination of the photon energy turned out to be the most important step. Two favorable energies were identified that correspond to two different scenarios---three-photon ionization of the $3s$ orbital simultaneously  with two-photon ionization of the $3p_0$ orbital for pulses with sufficiently large spectral bandwidth to render the photoelectrons energetically indistinguishable and one-photon ionization of the  $3p_0$ orbital combined with transitions between $3s$ and $3p_0$. 

When only the hole coherence is optimized, without any restriction on
the hole population, the population of the $3p_0$ hole is found to
exceed that of the $3s$ hole by two orders of magnitude or more. We have therefore extended the optimization functional to include a term that prescribes the population ratio. An equal or similar population of both hole states would be required when using the hole state superposition in time-dependent spectroscopy. SPA-optimization has addressed also this more challenging control task very successfully, yielding hole coherences close to one for exactly the population ratio desired, no matter whether the population of the $3s$ hole should exceed that of the $3p_0$ or vice versa or whether the populations should be equal. The resulting pulse shapes were found to be fairly simple, with their spectra indicating the second control scenario to be at work. 

In all optimizations for hole creation in argon, channel coupling was found not to play any role. This is in contrast to photoionization in xenon where channel coupling is the main source of decoherence~\cite{PabstPRL2011}. It may explain why, for argon, hole coherences very close to the absolute maximum can be achieved. Of course, this raises the question as to what the maximum hole coherence is in a case where channel coupling is known to be important. SPA-optimization is an ideal tool to address this question. 

\begin{acknowledgments}
  Financial support by the State Hessen Initiative for the
  Development of Scientific and Economic Excellence (LOEWE) within the
  focus project Electron Dynamic of Chiral Systems (ELCH) is                
  gratefully acknowledged. A.~K. is supported by the Louise-Johnson Fellowship of the Hamburg Centre for Ultrafast Imaging. 
\end{acknowledgments}

\bibliography{brent}

\begin{thebibliography}{36}%
\makeatletter
\providecommand \@ifxundefined [1]{%
 \@ifx{#1\undefined}
}%
\providecommand \@ifnum [1]{%
 \ifnum #1\expandafter \@firstoftwo
 \else \expandafter \@secondoftwo
 \fi
}%
\providecommand \@ifx [1]{%
 \ifx #1\expandafter \@firstoftwo
 \else \expandafter \@secondoftwo
 \fi
}%
\providecommand \natexlab [1]{#1}%
\providecommand \enquote  [1]{``#1''}%
\providecommand \bibnamefont  [1]{#1}%
\providecommand \bibfnamefont [1]{#1}%
\providecommand \citenamefont [1]{#1}%
\providecommand \href@noop [0]{\@secondoftwo}%
\providecommand \href [0]{\begingroup \@sanitize@url \@href}%
\providecommand \@href[1]{\@@startlink{#1}\@@href}%
\providecommand \@@href[1]{\endgroup#1\@@endlink}%
\providecommand \@sanitize@url [0]{\catcode `\\12\catcode `\$12\catcode
  `\&12\catcode `\#12\catcode `\^12\catcode `\_12\catcode `\%12\relax}%
\providecommand \@@startlink[1]{}%
\providecommand \@@endlink[0]{}%
\providecommand \url  [0]{\begingroup\@sanitize@url \@url }%
\providecommand \@url [1]{\endgroup\@href {#1}{\urlprefix }}%
\providecommand \urlprefix  [0]{URL }%
\providecommand \Eprint [0]{\href }%
\providecommand \doibase [0]{http://dx.doi.org/}%
\providecommand \selectlanguage [0]{\@gobble}%
\providecommand \bibinfo  [0]{\@secondoftwo}%
\providecommand \bibfield  [0]{\@secondoftwo}%
\providecommand \translation [1]{[#1]}%
\providecommand \BibitemOpen [0]{}%
\providecommand \bibitemStop [0]{}%
\providecommand \bibitemNoStop [0]{.\EOS\space}%
\providecommand \EOS [0]{\spacefactor3000\relax}%
\providecommand \BibitemShut  [1]{\csname bibitem#1\endcsname}%
\let\auto@bib@innerbib\@empty
\bibitem [{\citenamefont {Glaser}\ \emph {et~al.}(2015)\citenamefont {Glaser},
  \citenamefont {Boscain}, \citenamefont {Calarco}, \citenamefont {Koch},
  \citenamefont {K\"ockenberger}, \citenamefont {Kosloff}, \citenamefont
  {Kuprov}, \citenamefont {Luy}, \citenamefont {Schirmer}, \citenamefont
  {Schulte-Herbr\"uggen}, \citenamefont {Sugny},\ and\ \citenamefont
  {Wilhelm}}]{GlaserEPJD15}%
  \BibitemOpen
  \bibfield  {author} {\bibinfo {author} {\bibfnamefont {S.~J.}\ \bibnamefont
  {Glaser}}, \bibinfo {author} {\bibfnamefont {U.}~\bibnamefont {Boscain}},
  \bibinfo {author} {\bibfnamefont {T.}~\bibnamefont {Calarco}}, \bibinfo
  {author} {\bibfnamefont {C.~P.}\ \bibnamefont {Koch}}, \bibinfo {author}
  {\bibfnamefont {W.}~\bibnamefont {K\"ockenberger}}, \bibinfo {author}
  {\bibfnamefont {R.}~\bibnamefont {Kosloff}}, \bibinfo {author} {\bibfnamefont
  {I.}~\bibnamefont {Kuprov}}, \bibinfo {author} {\bibfnamefont
  {B.}~\bibnamefont {Luy}}, \bibinfo {author} {\bibfnamefont {S.}~\bibnamefont
  {Schirmer}}, \bibinfo {author} {\bibfnamefont {T.}~\bibnamefont
  {Schulte-Herbr\"uggen}}, \bibinfo {author} {\bibfnamefont {D.}~\bibnamefont
  {Sugny}}, \ and\ \bibinfo {author} {\bibfnamefont {F.~K.}\ \bibnamefont
  {Wilhelm}},\ }\href {\doibase 10.1140/epjd/e2015-60464-1} {\bibfield
  {journal} {\bibinfo  {journal} {Eur. Phys. J. D}\ }\textbf {\bibinfo {volume}
  {69}},\ \bibinfo {pages} {279} (\bibinfo {year} {2015})}\BibitemShut
  {NoStop}%
\bibitem [{\citenamefont {Klamroth}(2006)}]{KlamrothJCP06}%
  \BibitemOpen
  \bibfield  {author} {\bibinfo {author} {\bibfnamefont {T.}~\bibnamefont
  {Klamroth}},\ }\href {\doibase http://dx.doi.org/10.1063/1.2185633}
  {\bibfield  {journal} {\bibinfo  {journal} {J. Chem. Phys.}\ }\textbf
  {\bibinfo {volume} {124}},\ \bibinfo {eid} {144310} (\bibinfo {year}
  {2006})}\BibitemShut {NoStop}%
\bibitem [{\citenamefont {Greenman}\ \emph {et~al.}(2015)\citenamefont
  {Greenman}, \citenamefont {Koch},\ and\ \citenamefont
  {Whaley}}]{GreenmanPRA15}%
  \BibitemOpen
  \bibfield  {author} {\bibinfo {author} {\bibfnamefont {L.}~\bibnamefont
  {Greenman}}, \bibinfo {author} {\bibfnamefont {C.~P.}\ \bibnamefont {Koch}},
  \ and\ \bibinfo {author} {\bibfnamefont {K.~B.}\ \bibnamefont {Whaley}},\
  }\href {\doibase 10.1103/PhysRevA.92.013407} {\bibfield  {journal} {\bibinfo
  {journal} {Phys. Rev. A}\ }\textbf {\bibinfo {volume} {92}},\ \bibinfo
  {pages} {013407} (\bibinfo {year} {2015})}\BibitemShut {NoStop}%
\bibitem [{\citenamefont {Goetz}\ \emph {et~al.}(2016)\citenamefont {Goetz},
  \citenamefont {Karamatskou}, \citenamefont {Santra},\ and\ \citenamefont
  {Koch}}]{GoetzPRA16}%
  \BibitemOpen
  \bibfield  {author} {\bibinfo {author} {\bibfnamefont {R.~E.}\ \bibnamefont
  {Goetz}}, \bibinfo {author} {\bibfnamefont {A.}~\bibnamefont {Karamatskou}},
  \bibinfo {author} {\bibfnamefont {R.}~\bibnamefont {Santra}}, \ and\ \bibinfo
  {author} {\bibfnamefont {C.~P.}\ \bibnamefont {Koch}},\ }\href {\doibase
  10.1103/PhysRevA.93.013413} {\bibfield  {journal} {\bibinfo  {journal} {Phys.
  Rev. A}\ }\textbf {\bibinfo {volume} {93}},\ \bibinfo {pages} {013413}
  (\bibinfo {year} {2016})}\BibitemShut {NoStop}%
\bibitem [{\citenamefont {Mundt}\ and\ \citenamefont
  {Tannor}(2009)}]{MundtNJP09}%
  \BibitemOpen
  \bibfield  {author} {\bibinfo {author} {\bibfnamefont {M.}~\bibnamefont
  {Mundt}}\ and\ \bibinfo {author} {\bibfnamefont {D.~J.}\ \bibnamefont
  {Tannor}},\ }\href@noop {} {\bibfield  {journal} {\bibinfo  {journal} {New J.
  Phys.}\ }\textbf {\bibinfo {volume} {11}},\ \bibinfo {pages} {105038}
  (\bibinfo {year} {2009})}\BibitemShut {NoStop}%
\bibitem [{\citenamefont {Castro}\ \emph {et~al.}(2012)\citenamefont {Castro},
  \citenamefont {Werschnik},\ and\ \citenamefont {Gross}}]{CastroPRL12}%
  \BibitemOpen
  \bibfield  {author} {\bibinfo {author} {\bibfnamefont {A.}~\bibnamefont
  {Castro}}, \bibinfo {author} {\bibfnamefont {J.}~\bibnamefont {Werschnik}}, \
  and\ \bibinfo {author} {\bibfnamefont {E.~K.~U.}\ \bibnamefont {Gross}},\
  }\href {\doibase 10.1103/PhysRevLett.109.153603} {\bibfield  {journal}
  {\bibinfo  {journal} {Phys. Rev. Lett.}\ }\textbf {\bibinfo {volume} {109}},\
  \bibinfo {pages} {153603} (\bibinfo {year} {2012})}\BibitemShut {NoStop}%
\bibitem [{\citenamefont {Hellgren}\ \emph {et~al.}(2013)\citenamefont
  {Hellgren}, \citenamefont {R\"as\"anen},\ and\ \citenamefont
  {Gross}}]{HellgrenPRA13}%
  \BibitemOpen
  \bibfield  {author} {\bibinfo {author} {\bibfnamefont {M.}~\bibnamefont
  {Hellgren}}, \bibinfo {author} {\bibfnamefont {E.}~\bibnamefont
  {R\"as\"anen}}, \ and\ \bibinfo {author} {\bibfnamefont {E.~K.~U.}\
  \bibnamefont {Gross}},\ }\href {\doibase 10.1103/PhysRevA.88.013414}
  {\bibfield  {journal} {\bibinfo  {journal} {Phys. Rev. A}\ }\textbf {\bibinfo
  {volume} {88}},\ \bibinfo {pages} {013414} (\bibinfo {year}
  {2013})}\BibitemShut {NoStop}%
\bibitem [{\citenamefont {Young}\ \emph {et~al.}(2006)\citenamefont {Young},
  \citenamefont {Arms}, \citenamefont {Dufresne}, \citenamefont {Dunford},
  \citenamefont {Ederer}, \citenamefont {H\"ohr}, \citenamefont {Kanter},
  \citenamefont {Kr\"assig}, \citenamefont {Landahl}, \citenamefont {Peterson},
  \citenamefont {Rudati}, \citenamefont {Santra},\ and\ \citenamefont
  {Southworth}}]{YoungPRL06}%
  \BibitemOpen
  \bibfield  {author} {\bibinfo {author} {\bibfnamefont {L.}~\bibnamefont
  {Young}}, \bibinfo {author} {\bibfnamefont {D.~A.}\ \bibnamefont {Arms}},
  \bibinfo {author} {\bibfnamefont {E.~M.}\ \bibnamefont {Dufresne}}, \bibinfo
  {author} {\bibfnamefont {R.~W.}\ \bibnamefont {Dunford}}, \bibinfo {author}
  {\bibfnamefont {D.~L.}\ \bibnamefont {Ederer}}, \bibinfo {author}
  {\bibfnamefont {C.}~\bibnamefont {H\"ohr}}, \bibinfo {author} {\bibfnamefont
  {E.~P.}\ \bibnamefont {Kanter}}, \bibinfo {author} {\bibfnamefont
  {B.}~\bibnamefont {Kr\"assig}}, \bibinfo {author} {\bibfnamefont {E.~C.}\
  \bibnamefont {Landahl}}, \bibinfo {author} {\bibfnamefont {E.~R.}\
  \bibnamefont {Peterson}}, \bibinfo {author} {\bibfnamefont {J.}~\bibnamefont
  {Rudati}}, \bibinfo {author} {\bibfnamefont {R.}~\bibnamefont {Santra}}, \
  and\ \bibinfo {author} {\bibfnamefont {S.~H.}\ \bibnamefont {Southworth}},\
  }\href {\doibase 10.1103/PhysRevLett.97.083601} {\bibfield  {journal}
  {\bibinfo  {journal} {Phys. Rev. Lett.}\ }\textbf {\bibinfo {volume} {97}},\
  \bibinfo {pages} {083601} (\bibinfo {year} {2006})}\BibitemShut {NoStop}%
\bibitem [{\citenamefont {Goulielmakis}\ \emph {et~al.}(2010)\citenamefont
  {Goulielmakis}, \citenamefont {Loh}, \citenamefont {Wirth}, \citenamefont
  {Santra}, \citenamefont {Rohringer}, \citenamefont {Yakovlev}, \citenamefont
  {Zherebtsov}, \citenamefont {Pfeifer}, \citenamefont {Azzeer}, \citenamefont
  {Kling}, \citenamefont {Leone},\ and\ \citenamefont
  {Krausz}}]{GoulielmakisNature10}%
  \BibitemOpen
  \bibfield  {author} {\bibinfo {author} {\bibfnamefont {E.}~\bibnamefont
  {Goulielmakis}}, \bibinfo {author} {\bibfnamefont {Z.-H.}\ \bibnamefont
  {Loh}}, \bibinfo {author} {\bibfnamefont {A.}~\bibnamefont {Wirth}}, \bibinfo
  {author} {\bibfnamefont {R.}~\bibnamefont {Santra}}, \bibinfo {author}
  {\bibfnamefont {N.}~\bibnamefont {Rohringer}}, \bibinfo {author}
  {\bibfnamefont {V.~S.}\ \bibnamefont {Yakovlev}}, \bibinfo {author}
  {\bibfnamefont {S.}~\bibnamefont {Zherebtsov}}, \bibinfo {author}
  {\bibfnamefont {T.}~\bibnamefont {Pfeifer}}, \bibinfo {author} {\bibfnamefont
  {A.~M.}\ \bibnamefont {Azzeer}}, \bibinfo {author} {\bibfnamefont {M.~F.}\
  \bibnamefont {Kling}}, \bibinfo {author} {\bibfnamefont {S.~R.}\ \bibnamefont
  {Leone}}, \ and\ \bibinfo {author} {\bibfnamefont {F.}~\bibnamefont
  {Krausz}},\ }\href {\doibase 10.1038/nature09212} {\bibfield  {journal}
  {\bibinfo  {journal} {Nature}\ }\textbf {\bibinfo {volume} {466}},\ \bibinfo
  {pages} {739} (\bibinfo {year} {2010})}\BibitemShut {NoStop}%
\bibitem [{\citenamefont {Wirth}\ \emph {et~al.}(2011)\citenamefont {Wirth},
  \citenamefont {Hassan}, \citenamefont {Grgura{\v s}}, \citenamefont {Gagnon},
  \citenamefont {Moulet}, \citenamefont {Luu}, \citenamefont {Pabst},
  \citenamefont {Santra}, \citenamefont {Alahmed}, \citenamefont {Azzeer},
  \citenamefont {Yakovlev}, \citenamefont {Pervak}, \citenamefont {Krausz},\
  and\ \citenamefont {Goulielmakis}}]{WirthSci11}%
  \BibitemOpen
  \bibfield  {author} {\bibinfo {author} {\bibfnamefont {A.}~\bibnamefont
  {Wirth}}, \bibinfo {author} {\bibfnamefont {M.~T.}\ \bibnamefont {Hassan}},
  \bibinfo {author} {\bibfnamefont {I.}~\bibnamefont {Grgura{\v s}}}, \bibinfo
  {author} {\bibfnamefont {J.}~\bibnamefont {Gagnon}}, \bibinfo {author}
  {\bibfnamefont {A.}~\bibnamefont {Moulet}}, \bibinfo {author} {\bibfnamefont
  {T.~T.}\ \bibnamefont {Luu}}, \bibinfo {author} {\bibfnamefont
  {S.}~\bibnamefont {Pabst}}, \bibinfo {author} {\bibfnamefont
  {R.}~\bibnamefont {Santra}}, \bibinfo {author} {\bibfnamefont {Z.~A.}\
  \bibnamefont {Alahmed}}, \bibinfo {author} {\bibfnamefont {A.~M.}\
  \bibnamefont {Azzeer}}, \bibinfo {author} {\bibfnamefont {V.~S.}\
  \bibnamefont {Yakovlev}}, \bibinfo {author} {\bibfnamefont {V.}~\bibnamefont
  {Pervak}}, \bibinfo {author} {\bibfnamefont {F.}~\bibnamefont {Krausz}}, \
  and\ \bibinfo {author} {\bibfnamefont {E.}~\bibnamefont {Goulielmakis}},\
  }\href {\doibase 10.1126/science.1210268} {\bibfield  {journal} {\bibinfo
  {journal} {Science}\ }\textbf {\bibinfo {volume} {334}},\ \bibinfo {pages}
  {195} (\bibinfo {year} {2011})}\BibitemShut {NoStop}%
\bibitem [{\citenamefont {Calegari}\ \emph {et~al.}(2014)\citenamefont
  {Calegari}, \citenamefont {Ayuso}, \citenamefont {Trabattoni}, \citenamefont
  {Belshaw}, \citenamefont {De~Camillis}, \citenamefont {Anumula},
  \citenamefont {Frassetto}, \citenamefont {Poletto}, \citenamefont {Palacios},
  \citenamefont {Decleva}, \citenamefont {Greenwood}, \citenamefont
  {Mart{\'\i}n},\ and\ \citenamefont {Nisoli}}]{CalegariSci14}%
  \BibitemOpen
  \bibfield  {author} {\bibinfo {author} {\bibfnamefont {F.}~\bibnamefont
  {Calegari}}, \bibinfo {author} {\bibfnamefont {D.}~\bibnamefont {Ayuso}},
  \bibinfo {author} {\bibfnamefont {A.}~\bibnamefont {Trabattoni}}, \bibinfo
  {author} {\bibfnamefont {L.}~\bibnamefont {Belshaw}}, \bibinfo {author}
  {\bibfnamefont {S.}~\bibnamefont {De~Camillis}}, \bibinfo {author}
  {\bibfnamefont {S.}~\bibnamefont {Anumula}}, \bibinfo {author} {\bibfnamefont
  {F.}~\bibnamefont {Frassetto}}, \bibinfo {author} {\bibfnamefont
  {L.}~\bibnamefont {Poletto}}, \bibinfo {author} {\bibfnamefont
  {A.}~\bibnamefont {Palacios}}, \bibinfo {author} {\bibfnamefont
  {P.}~\bibnamefont {Decleva}}, \bibinfo {author} {\bibfnamefont {J.~B.}\
  \bibnamefont {Greenwood}}, \bibinfo {author} {\bibfnamefont {F.}~\bibnamefont
  {Mart{\'\i}n}}, \ and\ \bibinfo {author} {\bibfnamefont {M.}~\bibnamefont
  {Nisoli}},\ }\href {\doibase 10.1126/science.1254061} {\bibfield  {journal}
  {\bibinfo  {journal} {Science}\ }\textbf {\bibinfo {volume} {346}},\ \bibinfo
  {pages} {336} (\bibinfo {year} {2014})}\BibitemShut {NoStop}%
\bibitem [{\citenamefont {Pabst}\ \emph {et~al.}(2011)\citenamefont {Pabst},
  \citenamefont {Greenman}, \citenamefont {Ho}, \citenamefont {Mazziotti},\
  and\ \citenamefont {Santra}}]{PabstPRL2011}%
  \BibitemOpen
  \bibfield  {author} {\bibinfo {author} {\bibfnamefont {S.}~\bibnamefont
  {Pabst}}, \bibinfo {author} {\bibfnamefont {L.}~\bibnamefont {Greenman}},
  \bibinfo {author} {\bibfnamefont {P.~J.}\ \bibnamefont {Ho}}, \bibinfo
  {author} {\bibfnamefont {D.~A.}\ \bibnamefont {Mazziotti}}, \ and\ \bibinfo
  {author} {\bibfnamefont {R.}~\bibnamefont {Santra}},\ }\href {\doibase
  10.1103/PhysRevLett.106.053003} {\bibfield  {journal} {\bibinfo  {journal}
  {Phys. Rev. Lett.}\ }\textbf {\bibinfo {volume} {106}},\ \bibinfo {pages}
  {053003} (\bibinfo {year} {2011})}\BibitemShut {NoStop}%
\bibitem [{\citenamefont {Rybak}\ \emph {et~al.}(2011)\citenamefont {Rybak},
  \citenamefont {Amaran}, \citenamefont {Levin}, \citenamefont {Tomza},
  \citenamefont {Moszynski}, \citenamefont {Kosloff}, \citenamefont {Koch},\
  and\ \citenamefont {Amitay}}]{RybakPRL11}%
  \BibitemOpen
  \bibfield  {author} {\bibinfo {author} {\bibfnamefont {L.}~\bibnamefont
  {Rybak}}, \bibinfo {author} {\bibfnamefont {S.}~\bibnamefont {Amaran}},
  \bibinfo {author} {\bibfnamefont {L.}~\bibnamefont {Levin}}, \bibinfo
  {author} {\bibfnamefont {M.}~\bibnamefont {Tomza}}, \bibinfo {author}
  {\bibfnamefont {R.}~\bibnamefont {Moszynski}}, \bibinfo {author}
  {\bibfnamefont {R.}~\bibnamefont {Kosloff}}, \bibinfo {author} {\bibfnamefont
  {C.~P.}\ \bibnamefont {Koch}}, \ and\ \bibinfo {author} {\bibfnamefont
  {Z.}~\bibnamefont {Amitay}},\ }\href@noop {} {\bibfield  {journal} {\bibinfo
  {journal} {Phys. Rev. Lett.}\ }\textbf {\bibinfo {volume} {107}},\ \bibinfo
  {pages} {273001} (\bibinfo {year} {2011})}\BibitemShut {NoStop}%
\bibitem [{\citenamefont {Santra}\ \emph {et~al.}(2006)\citenamefont {Santra},
  \citenamefont {Dunford},\ and\ \citenamefont {Young}}]{SantraPRA06}%
  \BibitemOpen
  \bibfield  {author} {\bibinfo {author} {\bibfnamefont {R.}~\bibnamefont
  {Santra}}, \bibinfo {author} {\bibfnamefont {R.~W.}\ \bibnamefont {Dunford}},
  \ and\ \bibinfo {author} {\bibfnamefont {L.}~\bibnamefont {Young}},\ }\href
  {\doibase 10.1103/PhysRevA.74.043403} {\bibfield  {journal} {\bibinfo
  {journal} {Phys. Rev. A}\ }\textbf {\bibinfo {volume} {74}},\ \bibinfo
  {pages} {043403} (\bibinfo {year} {2006})}\BibitemShut {NoStop}%
\bibitem [{\citenamefont {Doria}\ \emph {et~al.}(2011)\citenamefont {Doria},
  \citenamefont {Calarco},\ and\ \citenamefont {Montangero}}]{DoriaPRL11}%
  \BibitemOpen
  \bibfield  {author} {\bibinfo {author} {\bibfnamefont {P.}~\bibnamefont
  {Doria}}, \bibinfo {author} {\bibfnamefont {T.}~\bibnamefont {Calarco}}, \
  and\ \bibinfo {author} {\bibfnamefont {S.}~\bibnamefont {Montangero}},\
  }\href {\doibase 10.1103/PhysRevLett.106.190501} {\bibfield  {journal}
  {\bibinfo  {journal} {Phys. Rev. Lett.}\ }\textbf {\bibinfo {volume} {106}},\
  \bibinfo {pages} {190501} (\bibinfo {year} {2011})}\BibitemShut {NoStop}%
\bibitem [{\citenamefont {Caneva}\ \emph {et~al.}(2011)\citenamefont {Caneva},
  \citenamefont {Calarco},\ and\ \citenamefont {Montangero}}]{CanevaPRA11}%
  \BibitemOpen
  \bibfield  {author} {\bibinfo {author} {\bibfnamefont {T.}~\bibnamefont
  {Caneva}}, \bibinfo {author} {\bibfnamefont {T.}~\bibnamefont {Calarco}}, \
  and\ \bibinfo {author} {\bibfnamefont {S.}~\bibnamefont {Montangero}},\
  }\href {\doibase 10.1103/PhysRevA.84.022326} {\bibfield  {journal} {\bibinfo
  {journal} {Phys. Rev. A}\ }\textbf {\bibinfo {volume} {84}},\ \bibinfo
  {pages} {022326} (\bibinfo {year} {2011})}\BibitemShut {NoStop}%
\bibitem [{\citenamefont {Lagarias}\ \emph {et~al.}(1998)\citenamefont
  {Lagarias}, \citenamefont {Reeds}, \citenamefont {Wright},\ and\
  \citenamefont {Wright}}]{JeffreySIAM98}%
  \BibitemOpen
  \bibfield  {author} {\bibinfo {author} {\bibfnamefont {J.~C.}\ \bibnamefont
  {Lagarias}}, \bibinfo {author} {\bibfnamefont {J.~A.}\ \bibnamefont {Reeds}},
  \bibinfo {author} {\bibfnamefont {M.~H.}\ \bibnamefont {Wright}}, \ and\
  \bibinfo {author} {\bibfnamefont {P.~E.}\ \bibnamefont {Wright}},\ }\href
  {\doibase 10.1137/S1052623496303470} {\bibfield  {journal} {\bibinfo
  {journal} {SIAM Journal on Optimization}\ }\textbf {\bibinfo {volume} {9}},\
  \bibinfo {pages} {112} (\bibinfo {year} {1998})}\BibitemShut {NoStop}%
\bibitem [{\citenamefont {Brent}(1973)}]{Brent1973}%
  \BibitemOpen
  \bibfield  {author} {\bibinfo {author} {\bibfnamefont {R.~P.}\ \bibnamefont
  {Brent}},\ }\href@noop {} {\emph {\bibinfo {title} {Algorithms for
  minimization without derivatives}}},\ \bibinfo {edition} {1st}\ ed.\
  (\bibinfo  {publisher} {Prentice-Hall Inc.},\ \bibinfo {address} {Englewood
  Cliffs, Princeton, New Jersey},\ \bibinfo {year} {1973})\BibitemShut
  {NoStop}%
\bibitem [{\citenamefont {Greenman}\ \emph {et~al.}(2010)\citenamefont
  {Greenman}, \citenamefont {Ho}, \citenamefont {Pabst}, \citenamefont
  {Kamarchik}, \citenamefont {Mazziotti},\ and\ \citenamefont
  {Santra}}]{GreenmanPRA10}%
  \BibitemOpen
  \bibfield  {author} {\bibinfo {author} {\bibfnamefont {L.}~\bibnamefont
  {Greenman}}, \bibinfo {author} {\bibfnamefont {P.~J.}\ \bibnamefont {Ho}},
  \bibinfo {author} {\bibfnamefont {S.}~\bibnamefont {Pabst}}, \bibinfo
  {author} {\bibfnamefont {E.}~\bibnamefont {Kamarchik}}, \bibinfo {author}
  {\bibfnamefont {D.~A.}\ \bibnamefont {Mazziotti}}, \ and\ \bibinfo {author}
  {\bibfnamefont {R.}~\bibnamefont {Santra}},\ }\href@noop {} {\bibfield
  {journal} {\bibinfo  {journal} {Phys. Rev. A}\ }\textbf {\bibinfo {volume}
  {82}},\ \bibinfo {pages} {023406} (\bibinfo {year} {2010})}\BibitemShut
  {NoStop}%
\bibitem [{\citenamefont {Pabst}\ \emph {et~al.}(2014)\citenamefont {Pabst},
  \citenamefont {Greenman},\ and\ \citenamefont {Santra}}]{xcid}%
  \BibitemOpen
  \bibfield  {author} {\bibinfo {author} {\bibfnamefont {S.}~\bibnamefont
  {Pabst}}, \bibinfo {author} {\bibfnamefont {L.}~\bibnamefont {Greenman}}, \
  and\ \bibinfo {author} {\bibfnamefont {R.}~\bibnamefont {Santra}},\
  }\href@noop {} {\enquote {\bibinfo {title} {{\textsc{XCID}} program package
  for multichannel ionization dynamics},}\ } (\bibinfo {year} {Rev 1425,
  2014})\BibitemShut {NoStop}%
\bibitem [{\citenamefont {Karamatskou}\ \emph {et~al.}(2014)\citenamefont
  {Karamatskou}, \citenamefont {Pabst}, \citenamefont {Chen},\ and\
  \citenamefont {Santra}}]{AntoniaPRA14}%
  \BibitemOpen
  \bibfield  {author} {\bibinfo {author} {\bibfnamefont {A.}~\bibnamefont
  {Karamatskou}}, \bibinfo {author} {\bibfnamefont {S.}~\bibnamefont {Pabst}},
  \bibinfo {author} {\bibfnamefont {Y.-J.}\ \bibnamefont {Chen}}, \ and\
  \bibinfo {author} {\bibfnamefont {R.}~\bibnamefont {Santra}},\ }\href
  {\doibase 10.1103/PhysRevA.89.033415} {\bibfield  {journal} {\bibinfo
  {journal} {Phys. Rev. A}\ }\textbf {\bibinfo {volume} {89}},\ \bibinfo
  {pages} {033415} (\bibinfo {year} {2014})}\BibitemShut {NoStop}%
\bibitem [{\citenamefont {Koopmans}(1934)}]{Koopmans1934}%
  \BibitemOpen
  \bibfield  {author} {\bibinfo {author} {\bibfnamefont {T.}~\bibnamefont
  {Koopmans}},\ }\href {\doibase
  http://dx.doi.org/10.1016/S0031-8914(34)90011-2} {\bibfield  {journal}
  {\bibinfo  {journal} {Physica}\ }\textbf {\bibinfo {volume} {1}},\ \bibinfo
  {pages} {104 } (\bibinfo {year} {1934})}\BibitemShut {NoStop}%
\bibitem [{\citenamefont {Grant}(2007)}]{Grand2007relativistic}%
  \BibitemOpen
  \bibfield  {author} {\bibinfo {author} {\bibfnamefont {I.~P.}\ \bibnamefont
  {Grant}},\ }\href@noop {} {\emph {\bibinfo {title} {Relativistic quantum
  theory of atoms and molecules : theory and computation}}},\ \bibinfo {series}
  {Springer series on atomic, optical, and plasma physics}, Vol.~\bibinfo
  {volume} {40}\ (\bibinfo  {publisher} {Springer},\ \bibinfo {address} {New
  York},\ \bibinfo {year} {2007})\BibitemShut {NoStop}%
\bibitem [{\citenamefont {Santra}\ and\ \citenamefont
  {Cederbaum}(2002)}]{SantraPhysRep02}%
  \BibitemOpen
  \bibfield  {author} {\bibinfo {author} {\bibfnamefont {R.}~\bibnamefont
  {Santra}}\ and\ \bibinfo {author} {\bibfnamefont {L.~S.}\ \bibnamefont
  {Cederbaum}},\ }\href {\doibase
  http://dx.doi.org/10.1016/S0370-1573(02)00143-6} {\bibfield  {journal}
  {\bibinfo  {journal} {Phys. Rep.}\ }\textbf {\bibinfo {volume} {368}},\
  \bibinfo {pages} {1 } (\bibinfo {year} {2002})}\BibitemShut {NoStop}%
\bibitem [{\citenamefont {Muga}\ \emph {et~al.}(2004)\citenamefont {Muga},
  \citenamefont {Palao}, \citenamefont {Navarro},\ and\ \citenamefont
  {Egusquiza}}]{MugaPhysRep04}%
  \BibitemOpen
  \bibfield  {author} {\bibinfo {author} {\bibfnamefont {J.~G.}\ \bibnamefont
  {Muga}}, \bibinfo {author} {\bibfnamefont {J.~P.}\ \bibnamefont {Palao}},
  \bibinfo {author} {\bibfnamefont {B.}~\bibnamefont {Navarro}}, \ and\
  \bibinfo {author} {\bibfnamefont {I.~L.}\ \bibnamefont {Egusquiza}},\ }\href
  {\doibase http://dx.doi.org/10.1016/j.physrep.2004.03.002} {\bibfield
  {journal} {\bibinfo  {journal} {Physics Reports}\ }\textbf {\bibinfo {volume}
  {395}},\ \bibinfo {pages} {357 } (\bibinfo {year} {2004})}\BibitemShut
  {NoStop}%
\bibitem [{\citenamefont {Riss}\ and\ \citenamefont {Meyer}(1993)}]{RissJPB93}%
  \BibitemOpen
  \bibfield  {author} {\bibinfo {author} {\bibfnamefont {U.~V.}\ \bibnamefont
  {Riss}}\ and\ \bibinfo {author} {\bibfnamefont {H.~D.}\ \bibnamefont
  {Meyer}},\ }\href@noop {} {\bibfield  {journal} {\bibinfo  {journal} {J.
  Phys. B: Atomic, Molecular and Optical Physics}\ }\textbf {\bibinfo {volume}
  {26}},\ \bibinfo {pages} {4503} (\bibinfo {year} {1993})}\BibitemShut
  {NoStop}%
\bibitem [{\citenamefont {Jolicard}\ and\ \citenamefont
  {Austin}(1985)}]{JolicardChemPhysLett85}%
  \BibitemOpen
  \bibfield  {author} {\bibinfo {author} {\bibfnamefont {G.}~\bibnamefont
  {Jolicard}}\ and\ \bibinfo {author} {\bibfnamefont {E.~J.}\ \bibnamefont
  {Austin}},\ }\href {\doibase http://dx.doi.org/10.1016/0009-2614(85)87164-5}
  {\bibfield  {journal} {\bibinfo  {journal} {Chem. Phys. Lett.}\ }\textbf
  {\bibinfo {volume} {121}},\ \bibinfo {pages} {106 } (\bibinfo {year}
  {1985})}\BibitemShut {NoStop}%
\bibitem [{\citenamefont {Rohringer}\ and\ \citenamefont
  {Santra}(2009)}]{RohringerPRA09}%
  \BibitemOpen
  \bibfield  {author} {\bibinfo {author} {\bibfnamefont {N.}~\bibnamefont
  {Rohringer}}\ and\ \bibinfo {author} {\bibfnamefont {R.}~\bibnamefont
  {Santra}},\ }\href {\doibase 10.1103/PhysRevA.79.053402} {\bibfield
  {journal} {\bibinfo  {journal} {Phys. Rev. A}\ }\textbf {\bibinfo {volume}
  {79}},\ \bibinfo {pages} {053402} (\bibinfo {year} {2009})}\BibitemShut
  {NoStop}%
\bibitem [{\citenamefont {Palao}\ \emph {et~al.}(2013)\citenamefont {Palao},
  \citenamefont {Reich},\ and\ \citenamefont {Koch}}]{PalaoPRA13}%
  \BibitemOpen
  \bibfield  {author} {\bibinfo {author} {\bibfnamefont {J.~P.}\ \bibnamefont
  {Palao}}, \bibinfo {author} {\bibfnamefont {D.~M.}\ \bibnamefont {Reich}}, \
  and\ \bibinfo {author} {\bibfnamefont {C.~P.}\ \bibnamefont {Koch}},\ }\href
  {\doibase 10.1103/PhysRevA.88.053409} {\bibfield  {journal} {\bibinfo
  {journal} {Phys. Rev. A}\ }\textbf {\bibinfo {volume} {88}},\ \bibinfo
  {pages} {053409} (\bibinfo {year} {2013})}\BibitemShut {NoStop}%
\bibitem [{\citenamefont {Reich}\ \emph {et~al.}(2014)\citenamefont {Reich},
  \citenamefont {Palao},\ and\ \citenamefont {Koch}}]{ReichJMO14}%
  \BibitemOpen
  \bibfield  {author} {\bibinfo {author} {\bibfnamefont {D.~M.}\ \bibnamefont
  {Reich}}, \bibinfo {author} {\bibfnamefont {J.~P.}\ \bibnamefont {Palao}}, \
  and\ \bibinfo {author} {\bibfnamefont {C.~P.}\ \bibnamefont {Koch}},\ }\href
  {\doibase 10.1080/09500340.2013.844866} {\bibfield  {journal} {\bibinfo
  {journal} {J. Mod. Opt.}\ }\textbf {\bibinfo {volume} {61}},\ \bibinfo
  {pages} {822} (\bibinfo {year} {2014})}\BibitemShut {NoStop}%
\bibitem [{\citenamefont {Reich}\ \emph {et~al.}(2012)\citenamefont {Reich},
  \citenamefont {Ndong},\ and\ \citenamefont {Koch}}]{ReichJCP12}%
  \BibitemOpen
  \bibfield  {author} {\bibinfo {author} {\bibfnamefont {D.~M.}\ \bibnamefont
  {Reich}}, \bibinfo {author} {\bibfnamefont {M.}~\bibnamefont {Ndong}}, \ and\
  \bibinfo {author} {\bibfnamefont {C.~P.}\ \bibnamefont {Koch}},\ }\href
  {\doibase 10.1063/1.3691827} {\bibfield  {journal} {\bibinfo  {journal} {J.
  Chem. Phys.}\ }\textbf {\bibinfo {volume} {136}},\ \bibinfo {pages} {104103}
  (\bibinfo {year} {2012})}\BibitemShut {NoStop}%
\bibitem [{\citenamefont {Ndong}\ \emph {et~al.}(2009)\citenamefont {Ndong},
  \citenamefont {Tal-Ezer}, \citenamefont {Kosloff},\ and\ \citenamefont
  {Koch}}]{NdongJCP09}%
  \BibitemOpen
  \bibfield  {author} {\bibinfo {author} {\bibfnamefont {M.}~\bibnamefont
  {Ndong}}, \bibinfo {author} {\bibfnamefont {H.}~\bibnamefont {Tal-Ezer}},
  \bibinfo {author} {\bibfnamefont {R.}~\bibnamefont {Kosloff}}, \ and\
  \bibinfo {author} {\bibfnamefont {C.~P.}\ \bibnamefont {Koch}},\ }\href
  {\doibase 10.1063/1.3098940} {\bibfield  {journal} {\bibinfo  {journal} {J.
  Chem. Phys.}\ }\textbf {\bibinfo {volume} {130}},\ \bibinfo {pages} {124108}
  (\bibinfo {year} {2009})}\BibitemShut {NoStop}%
\bibitem [{\citenamefont {Nelder}\ and\ \citenamefont
  {Mead}(1964)}]{NelderMead}%
  \BibitemOpen
  \bibfield  {author} {\bibinfo {author} {\bibfnamefont {J.~A.}\ \bibnamefont
  {Nelder}}\ and\ \bibinfo {author} {\bibfnamefont {R.}~\bibnamefont {Mead}},\
  }\href@noop {} {\bibfield  {journal} {\bibinfo  {journal} {The Computer
  Journal}\ }\textbf {\bibinfo {volume} {7}},\ \bibinfo {pages} {308} (\bibinfo
  {year} {1964})}\BibitemShut {NoStop}%
\bibitem [{\citenamefont {Klein}\ and\ \citenamefont
  {Neira}(2013)}]{Klein2013}%
  \BibitemOpen
  \bibfield  {author} {\bibinfo {author} {\bibfnamefont {K.}~\bibnamefont
  {Klein}}\ and\ \bibinfo {author} {\bibfnamefont {J.}~\bibnamefont {Neira}},\
  }\href {\doibase 10.1007/s10614-013-9377-8} {\bibfield  {journal} {\bibinfo
  {journal} {Computational Economics}\ }\textbf {\bibinfo {volume} {43}},\
  \bibinfo {pages} {447} (\bibinfo {year} {2013})}\BibitemShut {NoStop}%
\bibitem [{\citenamefont {Rach}\ \emph {et~al.}(2015)\citenamefont {Rach},
  \citenamefont {M\"uller}, \citenamefont {Calarco},\ and\ \citenamefont
  {Montangero}}]{RachPRA15}%
  \BibitemOpen
  \bibfield  {author} {\bibinfo {author} {\bibfnamefont {N.}~\bibnamefont
  {Rach}}, \bibinfo {author} {\bibfnamefont {M.~M.}\ \bibnamefont {M\"uller}},
  \bibinfo {author} {\bibfnamefont {T.}~\bibnamefont {Calarco}}, \ and\
  \bibinfo {author} {\bibfnamefont {S.}~\bibnamefont {Montangero}},\ }\href
  {\doibase 10.1103/PhysRevA.92.062343} {\bibfield  {journal} {\bibinfo
  {journal} {Phys. Rev. A}\ }\textbf {\bibinfo {volume} {92}},\ \bibinfo
  {pages} {062343} (\bibinfo {year} {2015})}\BibitemShut {NoStop}%
\bibitem [{\citenamefont {Goerz}\ \emph {et~al.}(2015)\citenamefont {Goerz},
  \citenamefont {Whaley},\ and\ \citenamefont {Koch}}]{GoerzEPJQT15}%
  \BibitemOpen
  \bibfield  {author} {\bibinfo {author} {\bibfnamefont {M.~H.}\ \bibnamefont
  {Goerz}}, \bibinfo {author} {\bibfnamefont {K.~B.}\ \bibnamefont {Whaley}}, \
  and\ \bibinfo {author} {\bibfnamefont {C.~P.}\ \bibnamefont {Koch}},\ }\href
  {\doibase 10.1140/epjqt/s40507-015-0034-0} {\bibfield  {journal} {\bibinfo
  {journal} {EPJ Quantum Technology}\ }\textbf {\bibinfo {volume} {2}},\
  \bibinfo {pages} {21} (\bibinfo {year} {2015})}\BibitemShut {NoStop}%
\end{thebibliography}%

\end{document}